\documentclass[journal]{IEEEtran}
\usepackage{xcolor,soul,framed} 
\colorlet{shadecolor}{yellow}
\usepackage[pdftex]{graphicx}
\graphicspath{{../pdf/}{../jpeg/}}
\DeclareGraphicsExtensions{.pdf,.jpeg,.png}
\usepackage[cmex10]{amsmath}
\usepackage{array}
\usepackage{algorithmic}
\usepackage{algorithm}
\usepackage{amsfonts}
\usepackage{amsmath}
\usepackage{bm}
\usepackage{bbm,dsfont}
\usepackage{booktabs}
\usepackage{blindtext}
\usepackage[colorinlistoftodos]{todonotes}
\usepackage{xcolor}
\usepackage{comment}
\usepackage{amsthm}
\usepackage{cite}
\usepackage{subfigure}
\usepackage{enumerate}
\usepackage{enumitem}
\usepackage{eqparbox}
\usepackage[T1]{fontenc}
\usepackage{flushend}
\usepackage{footnote}
\usepackage{graphicx}
\usepackage{makecell}
\usepackage{mathtools}
\usepackage{mhchem}
\usepackage{mdwmath}
\usepackage{mdwtab}
\usepackage{multirow}
\usepackage{nomencl}
\usepackage{stfloats}
\usepackage{textcase}
\usepackage{threeparttable}
\usepackage{url}
\usepackage{setspace}

\usepackage{etoolbox}
\makeatletter
\patchcmd{\@makecaption}
  {\addvspace{0.4\baselineskip}\egroup}
  {\addvspace{-1\baselineskip}\egroup}
  {}
  {}
\makeatother
\begin{document}

\title{\vspace{-0.48cm}\!Transient Stability-Driven Planning for the Optimal Sizing of Resilient AC/DC Hybrid Microgrids}
\IEEEaftertitletext{\vspace{-2.08\baselineskip}}

\author{Yi Wang,~\IEEEmembership{Member,~IEEE,}
        and Goran Strbac,~\IEEEmembership{Member,~IEEE}

\thanks{This work was supported by the UK EPSRC project: `Integrated Development of Low-Carbon Energy Systems (IDLES): A Whole-System Paradigm for Creating a National Strategy' (project code: EP/R045518/1) and the Horizon Europe project: `Reliability, Resilience and Defense technology for the griD' (Grant agreement ID: 101075714).}
}

\markboth{IEEE Trans. Power Syst., Accepted for Publication}%
{Shell \MakeLowercase{\textit{et al.}}: Bare Demo of IEEEtran.cls for IEEE Journals}
\maketitle

\begin{abstract}
This paper proposes a transient stability-driven planning framework for the optimal sizing problem of resilient AC/DC hybrid microgrids (HMGs) under different types of contingencies, capturing frequency and voltage stability requirements as well as the frequency-voltage coupling dynamics of AC/DC interlinking converters (ICs). The planning model is formulated into a defender-attacker-defender (DAD) architecture, which can be further merged into two levels, i.e., upper-level and low-level problems, and then iteratively solved by an enhanced genetic algorithm with sparsity calculation and local search. Regarding the operation stage, a novel transient stability-constrained optimal power flow (TSC-OPF) algorithm is proposed for static and transient operations of HMGs, capturing governor dynamics and automatic voltage regulator of conventional generators as well as the droop control dynamics of inverter-based resources (IBRs) for frequency control and voltage control, respectively. Furthermore, a Lyapunov optimisation approach is developed to capture the time-coupling property of energy storages (ESs) and then allow the TSC-OPF to be solved on an hourly basis with a second-scale resolution, achieving the co-optimisation of static and transient stability requirements. Case studies have been conducted to verify the effectiveness of the proposed planning framework in obtaining cost-effective investment decisions for various resources while respecting transient stability requirements under different contingencies.
\end{abstract}
\vspace{-0.28em}
\vspace{-0.28em}

\renewcommand{\nomgroup}[1]{%
\ifthenelse{\equal{#1}{A}}{\item[\emph{A.~Indices~and~Sets}]}{%
\ifthenelse{\equal{#1}{B}}{\item[\emph{B.~Parameters}]}{%
\ifthenelse{\equal{#1}{C}}{\item[\emph{C.~Variables}]}{{}}}}
}
\makenomenclature
\setlength{\nomlabelwidth}{2.05cm} 
\setlength{\nomitemsep}{0.28mm} 
\nomenclature[A1]{$g \in \mathcal{DG}$}{Index and set of DGs}
\nomenclature[A2]{$g \in \mathcal{IBR}$}{Index and set of IBRs}
\nomenclature[A3]{$k \in \mathcal{ES}$}{Index and set of ESs}
\nomenclature[A4]{$b \in \mathcal{B}$}{Index and set of buses}
\nomenclature[A5]{$i \in \mathcal{MG}^{ac}$}{Index and set of AC subgrids}
\nomenclature[A6]{$j \in \mathcal{MG}^{dc}$}{Index and set of DC subgrids}

\nomenclature[B01]{$\omega^{0}$}{Rotor's nominal speed}
\nomenclature[B02]{$H_{g}$}{Inertia constant of DG $g$}
\nomenclature[B03]{$X^{d}_{g}$}{Stator’s $d$ axis reactances of DG $g$}
\nomenclature[B04]{$X^{d'}_{g}$}{Transient reactance related to DG $g$}
\nomenclature[C01]{$\delta_{g}$}{Rotor angle of DG $g$}
\nomenclature[C02]{$\omega_{g}$}{Rotor's instantaneous speed of DG $g$}
\nomenclature[C03]{$f^{s}$}{System frequency}
\nomenclature[C04]{$i^{d}_{g},i^{q}_{g}$}{Stator's $d$ and $q$ axis currents}
\nomenclature[C06]{$v^{d}_{g},v^{q}_{g}$}{Stator's $d$ and $q$ axis voltages}
\nomenclature[C08]{$E^{q'}_{g}$}{Back-emf on the $q$ axis}
\nomenclature[C08]{$E^{fd}_{g}$}{Applied excitation voltage}
\nomenclature[C09]{$P^{m}_{g}$}{Applied mechanical power of turbine governor}
\nomenclature[C10]{$P^{e}_{g},Q^{e}_{g}$}{Electromagnetic/Reactive power of DG $g$}
\nomenclature[C11]{$P^{ibr}_{g},Q^{ibr}_{g}$}{Active/Reactive power of IBR $g$}
\nomenclature[C12]{$P_{ij}^{ic}$}{Power exchange between subgrids $i$ and $j$}
\nomenclature[C14]{$P_{k,t}^{c},P_{k,t}^{d}$}{Charging and discharging power of ES $k$}
\nomenclature[C15]{$E_{k,t}^{es}$}{Battery energy content of ES $k$}

\printnomenclature
\vspace{-0mm}

\section{Introduction}
\vspace{-0.00em}
\label{sec:I}
\IEEEPARstart{M}icrogrids (MGs) are playing a crucial role in the development of future energy systems worldwide, offering substantial advantages over traditional top-down radial network configurations \cite{li2017networked}. Despite great uncertainty surrounding the specific MG architectures that will emerge, there is clear evidence that building-scale and local-aggregator MGs will become more common in the coming decades. As smaller and localised power systems equipped with advanced control features and islanding capabilities, MGs are recognised as a viable option to encapsulate and coordinate a variety of distributed energy resources (DERs), such as photovoltaics (PVs), wind turbines (WTs), energy storage systems (ESs), etc., and can feature different configurations, e.g., AC, DC, and AC/DC hybrid MGs (HMGs) \cite{wang2020microgrids}. In particular, it has been demonstrated that HMGs surpass traditional AC MGs in terms of resilient response to transient emergencies and economic operations, due to their advantages of reducing multiple power conversions and incorporating DERs with DC power \cite{hussain2019microgrids}. In particular, AC and DC subgrids can effectively exchange power through AC/DC interlinking converters (ICs) to support transient stability \cite{li2017networked}. 

Prior research has extensively explored the planning and operational challenges of HMGs. In terms of operational issues, the field has been well-established, with numerous significant studies dedicated to this aspect, as highlighted by \cite{li2021distributed,zhang2023coordinated,pompodakis2020generic}. However, research on the planning problems of HMGs remains comparatively underdeveloped. In \cite{xie2022optimal}, a joint optimal sizing model for renewable energy resources (RESs) and ESs in HMGs is developed, considering the trade-off between operating economy and environmental conservation. In \cite{hamad2018optimal}, a branch-and-bound optimisation approach is developed for the optimal sizing of HMGs, considering RESs, ESs, and power converters. However, the above two papers overlook the influence of potential contingencies (e.g., sudden wind drops) on final investment decisions, which may be unrealistic. To address this issue, an AC/DC optimal expansion planning model considering topological constraints is proposed in \cite{de2023optimal} for the optimal sizing of converters; nevertheless, the sizing decisions of DERs are not considered. In \cite{rousis2019planning}, a two-stage planning model based on meta-heuristic techniques is proposed for the optimal sizing of HMGs. However, the investment decisions are deterministic and based on only one specific contingency (e.g., IC outage), which can not be generalised and adapted to different types of contingencies. In \cite{gong2021security}, a security-constrained planning model based on post-contingency corrective rescheduling is developed for the optimal sizing of ESs in an islanded HMG. However, frequency characteristics are considered in a static manner, which can not ensure secure HMG operations under transient scenarios.

Going further, it can be concluded that all the above planning models for HMGs \cite{xie2022optimal,hamad2018optimal,de2023optimal,rousis2019planning,gong2021security} only consider steady-state operations, while neglecting the influence of transient system dynamics on investment decisions, which can lead to impractical and insecure HMG operations. The increasing penetration of RESs such as WTs and PVs has caused the reduced inertia level of future power systems, which further increases the risk of transient frequency and voltage instability issues. For example, a blackout in South Australia witnessed the Rate of Change of Frequency (RoCoF) reaching 6.25 Hz/s and frequency dropping below 47 Hz, while a blackout in the UK occurred after the frequency dropped to 48.8 Hz \cite{li2021frequency}. In particular, unlike traditional bulk grids, MGs are inherently faced with a lack of rotational inertia affecting their security in the event of significant power imbalance. Without the consideration of transient stability, MGs might be at risk of devastating consequences such as under-frequency load shedding (UFLS), generator tripping, and blackout, where the UFLS cost can be as high as 1600-55000 USD/MWh \cite{o2021probabilistic}. Therefore, to adequately safeguard MG security and ensure sufficient operational capabilities, transient stability shall be considered from both planning and operation perspectives.

In recent years, there have been papers \cite{li2021frequency,paturet2020stochastic,badesa2022assigning,zhang2020modeling} deriving and incorporating frequency-related constraints into power system planning \cite{li2021frequency,paturet2020stochastic} and operation \cite{badesa2022assigning,zhang2020modeling} problems; nevertheless, the above papers are based on either simplified dynamic models \cite{li2021frequency,paturet2020stochastic} or linearised frequency constraints \cite{badesa2022assigning,zhang2020modeling} to simplify transient stability modelling. Furthermore, the simplifications therein are calculated from the characteristics of transmission systems rather than MGs or distribution systems. Regarding distribution level, a planning model based on stochastic-robust optimisation is developed in \cite{nakiganda2022stochastic} for resilient MGs, considering dynamic frequency response during islanding periods. In \cite{nakiganda2022resilient}, an iterative dynamic optimisation approach is proposed to incorporate transient and static constraints for the secure islanding operations of MGs. However, the iterative process in the above two papers \cite{nakiganda2022stochastic,nakiganda2022resilient} considers static models and transient simulations separately, which may lead to suboptimal solutions. In addition, these linearisation methods in \cite{nakiganda2022stochastic,nakiganda2022resilient} can only derive approximate constraints based on certain operation states, lacking the ability to generalise across diverse operating conditions. Furthermore, linearised power flow algorithms are used in \cite{nakiganda2022stochastic,nakiganda2022resilient} for static AC MG operations, which can only be applied for radial network topologies. However, other than traditional distribution networks, MGs might have different structures, e.g., meshed networks \cite{wang2021three}. Meshed networks include a more uniform power flow and can provide benefits for improving frequency and voltage profiles as well as reducing power loss, which introduce stronger capabilities to withstand contingencies and improve stability \cite{liu2018coordinated}.

In this context, transient stability-constrained optimal power flow (TSC-OPF) can be regarded as a suitable option to incorporate both detailed transient stability dynamics and non-linear power flow, formulating them as a set of differential-algebraic equations (DAEs). For instance, authors in \cite{xia2020transient,trivino2023network} propose TSC-OPF models to capture intricate frequency dynamics in a second scale, even though they still use iterative methods to solve TSC-OPF, which cannot guarantee optimal solutions. Addressing this limitation, a combinatorial TSC-OPF model is proposed in \cite{zhao2020frequency}, where frequency dynamics are directly embedded as DAEs within the optimisation process as equality constraints; thus, optimal dispatches can be obtained by solving the co-optimisation model only once. 
Despite the advancements in TSC-OPF models \cite{xia2020transient,trivino2023network,zhao2020frequency}, several critical limitations persist: 1) they primarily focus on frequency dynamics, while ignoring potential voltage instability issues; 2) these papers together with \cite{nakiganda2022stochastic,nakiganda2022resilient} all focus exclusively on conventional AC MGs rather than HMGs with ICs and frequency-voltage coupling features; 3) due to its highly non-linear nature, TSC-OPF is inherently time-consuming and more suitable for one-step operation problems with transient condition changes (second or minute scale) than long-term operation and planning problems with time-coupled features and static condition changes (daily scale). Thus, developing a comprehensive TSC-OPF algorithm and then applying it effectively to HMG planning problems can be very challenging.

To fill the above research gaps, this paper proposes a planning framework for the optimal sizing of HMGs with detailed transient stability dynamics and power flows. A novel TSC-OPF algorithm considering both frequency and voltage stability is developed for secure HMG operations, while the daily time-coupled optimisation problem is decoupled and shifted to an hourly basis via Lyapunov optimisation \cite{shi2022lyapunov}, thereby facilitating incorporation of TSC-OPF into the planning framework. Detailed contributions are summarised below:

\begin{enumerate}[label=\arabic*)]
\item Propose a stability-driven planning framework with a defender-attacker-defender (DAD) structure for the optimal sizing of resilient HMGs, capturing transient stability requirements under different types of contingencies, e.g., sudden RES drop, sudden load change, islanding, IC outage, etc.

\item Develop a novel TSC-OPF model considering both static operation constraints as well as frequency and voltage stability dynamics for the secure operation of HMGs with detailed IC dynamics. Regarding frequency stability, a synchronous machine model is introduced for generator dynamics, while frequency control with prime mover dynamics is incorporated based on PI control loops. Regarding voltage stability, automatic voltage regulator (AVR) of generators with a standard DC exciter and a stabilising feedback block as well as the voltage droop control dynamics of inverter-based resources (IBRs) are incorporated into the TSC-OPF model. In addition, a droop control scheme is introduced for the IC to link AC frequency and DC bus voltage for effective power exchange between AC subgrid and DC subgrid.

\item Introducing a Lyapunov optimisation approach with multiple weighting parameters to decouple the daily time-coupled static optimisation, allowing the developed TSC-OPF to be resolved on an hourly basis with a second-scale resolution. The weighting parameters are used for the trade-off between queue backlog reduction and immediate cost towards realistic optimisation results.

\item Reformulating the DAD planning model into a two-level structure solved by an iterative process. An enhanced genetic algorithm (GA) with adaptive crossover/mutation probabilities as well as sparsity calculation and local search is proposed to address low-level and upper-level problems, which can better avoid local optimum and speed up convergence.
\end{enumerate}

The rest of this paper is organised as follows. Section \ref{sec:II} presents the outline of the proposed stability-driven planning framework. Section \ref{sec:III} details the mathematical formulations of the planning model, while Section \ref{sec:IV} provides the solving procedure. In Section \ref{sec:V}, case studies are carried out and analysed under different contingencies. Section \ref{sec:VI} draws the conclusions and key findings of this paper.
\vspace{-0.58em}

\section{Outline of the proposed planning framework}
\label{sec:II}
\begin{figure}[h!] 
\vspace{-0.38em}
\centering  
\includegraphics[width=0.475\textwidth]{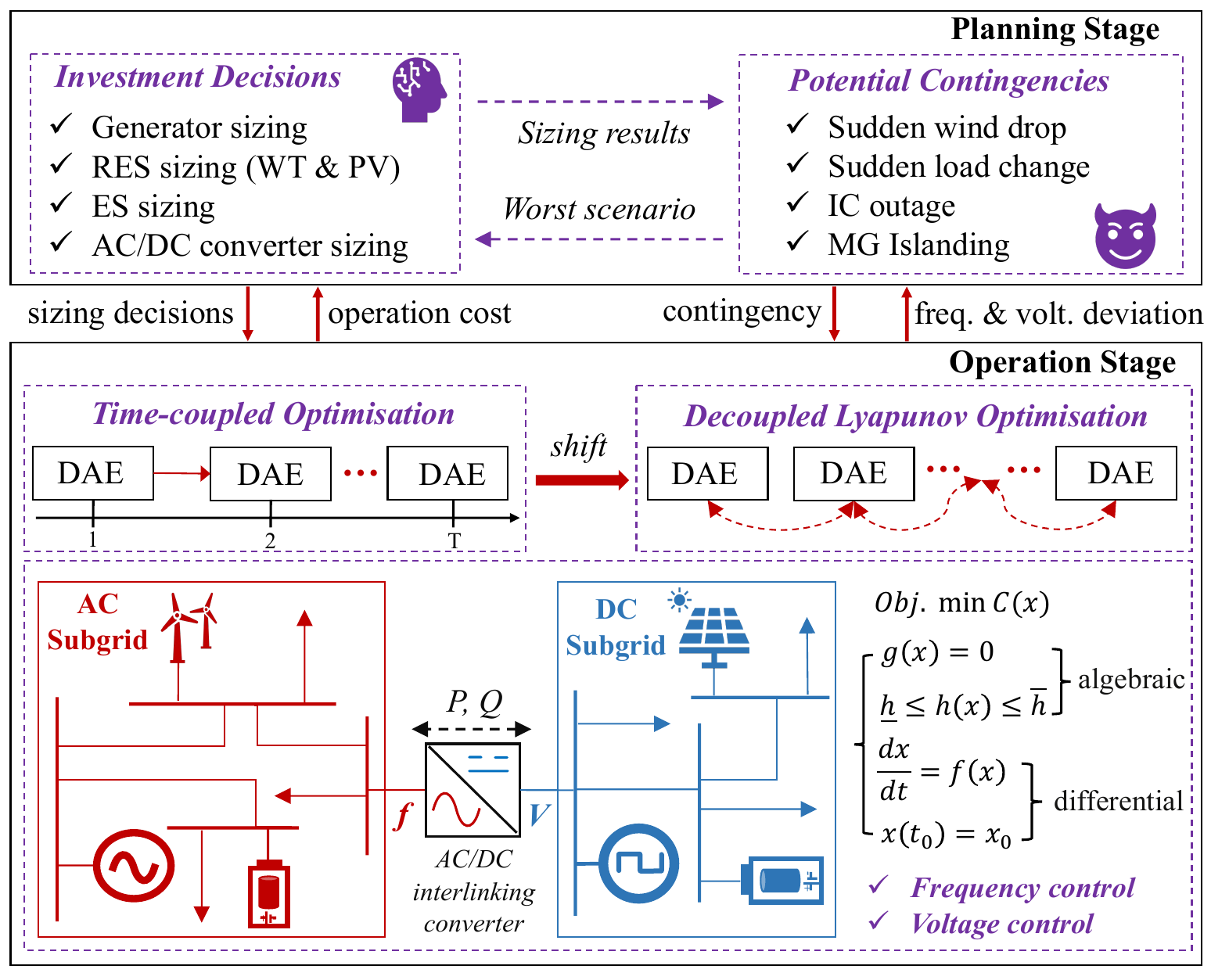}
\vspace{-0.58em}
\caption{Illustration of the proposed stability-driven planning model for the optimal sizing of AC/DC hybrid MGs.}
\label{fig:problem}
\vspace{-0.38em}
\end{figure}

This paper focuses on the transient stability-driven optimal sizing problem of HMGs, developing a novel planning framework for investment decisions of various DERs including conventional distributed generators (DGs), WTs, PVs and ESs that can maintain transient stability requirements for both frequency and voltage under various contingencies. 
In detail, unlike large transmission systems, this paper mainly focuses on the transient frequency stability and voltage stability problems of HMGs, corresponding to the ability of the HMG to withstand and stabilise after short-duration disturbances, e.g., sudden wind drop, load change, islanding, etc., following the similar practice in \cite{nakiganda2022resilient,nakiganda2022stochastic}. The studied HMG system is depicted in Fig. \ref{fig:problem}, where AC and DC subgrids can exchange power through frequency-voltage coupling features of ICs. On the AC side, electric components including DGs, WTs, ESs, and AC loads are located across different AC buses, while components including DGs, PVs, ESs, and DC loads are appropriately allocated on the DC side.

Specifically, at the planning stage, the proposed DAD model can be reformulated into two levels, i.e., upper-level problem and low-level problem \cite{wang2021three}. The low-level problem based on the attacker is adopted to simulate HMG operations under various types of contingencies such as sudden wind drop, load change, islanding and IC outages, and then select the worst-case scenario with the largest frequency and voltage deviations. The upper-level problem based on the defender is designed for the optimal sizing of the studied HMG against the worst-case scenario. On one hand, the low-level attacker model and upper-level defender model will be solved iteratively until transient stability requirements are met under any simulated contingencies. On the other hand, both the defender and attacker need to interact with the TSC-OPF operation model during the iterative process for contingency simulation and verification of investment decisions, respectively.

Regarding the operation stage, a novel TSC-OPF algorithm based on a set of DAEs is deployed for HMG operations, capturing the transient stability requirements of frequency and voltage as well as the frequency-voltage coupling of ICs. In detail, governor control dynamics and AVR of DGs and voltage droop control dynamics of IBRs have been incorporated into the TSC-OPF for frequency and voltage regulation, respectively. Furthermore, as depicted in Fig. \ref{fig:problem}, a Lyapunov optimisation approach is developed to handle the time-coupled features of ESs, transferring the HMG operation from a daily time-coupled fashion to a decoupled hourly solution. In this case, the TSC-OPF can be solved on an hourly basis with a second-scale resolution, allowing it to be appropriately incorporated into the DAD planning framework and achieve the co-optimisation of static and transient stability constraints.
\vspace{-1.4em}

\section{Mathematical Formulations of the Proposed Planning Framework}
\label{sec:III}
The proposed stability-driven planning model takes into account both transient dynamics and static operations of the HMG system. Overall, the model includes four types of constraints with details given below: 
\vspace{-0.7em}

\subsection{Generator Dynamics}
\label{sec:III.A}
\subsubsection{Conventional Diesel Generators}
A structure-preserving system formulation with a classical synchronous machine model \cite{machowski1997power} is introduced for DG dynamics. In detail, deviating from the reference frame speed, the rotor speed of DG $g$ can be expressed in the Laplace domain as
\begin{equation}\label{eq:rotor}
s \delta_{g} = \omega_{g} - \omega^0 = \Delta \omega_{g},\forall g \in \mathcal{DG},
\end{equation}
where $\delta_{g}$ refers to the rotor angle of DG $g$ expressed in $rad$. $\omega_{g}$ and $\omega^{0}$ are the rotor's instantaneous and nominal speed, while $\Delta \omega_g$ is the rotor's deviation from $\omega^{0}$. Then, the swing equation representing the rotor acceleration of DG $g$ is expressed as
\begin{equation}
s^2 \delta_{g} = s \Delta \omega_{g} = \omega^0 /2 H_{g} (P^{m}_{g} -P^{e}_{g}),\forall g \in \mathcal{DG},
\end{equation}
where $P^{m}_{g}$ is the applied mechanical power of the turbine governor expressed in $p.u.$. $P^{e}_{g}$ corresponds to the electromagnetic power of DG $g$, while $H_g$ is the inertia constant. Using the centre of inertia concept \cite{machowski1997power}, the system frequency $f^{s}$ of the HMG is defined as weighted average frequency, expressed as
\begin{equation}
f^{s} = \frac{\sum_{g \in \mathcal{DG}}\omega_{g}H_{g}}{2\pi\sum_{g \in \mathcal{DG}}H_{g}}.
\end{equation}

For simplicity, the resistance and stator sub-transient dynamics are ignored \cite{zhao2020frequency}; thus, the flux decay dynamics of DG $g$ can be expressed via the following set of stator equations:
\begin{equation}
v^{d}_{g} = X^{d'}_{g} i^q_{g},\forall g \in \mathcal{DG},
\end{equation}
\begin{equation}
E^{q'}_{g} = X^{d'}_{g} i^d_{g} + v^q_{g},\forall g \in \mathcal{DG},
\end{equation}
\begin{equation}
v^{d}_{g} = V_b sin(\delta_{g}-\delta_{b}),\forall g \in \mathcal{DG},\forall b \in \mathcal{B}_{dg},
\end{equation}
\begin{equation}
v^q_{g} = V_b cos(\delta_{g}-\delta_{b}),\forall g \in \mathcal{DG},\forall b \in \mathcal{B}_{dg},
\end{equation}
where $i^d_{g}$ and $i^q_{g}$ are the stator's $d$ and $q$ axis currents, while $v^d_{g}$ and $v^q_{g}$ denote the $d$ and $q$ axis voltages, respectively. $E_g^{q'}$ refers to the back-emf on the $q$ axis due to rotor field currents. $\mathcal{B}_{dg}$ refers to the bus set connected with DGs.

In addition, the rotor resistance and transient dynamics of DG $g$ can be fully modelled in the Laplace domain via the following circuit differential equation:
\begin{equation}
E^{fd}_{g} = E_{g}^{q'} + T^{d0'}_{g} s E_{g}^{q'} + (X^d_{g} - X^{d,'}_{g}) i^d_{g},\forall g \in \mathcal{DG},
\end{equation}
where $E^{fd}_{g}$ is the applied excitation voltage. $T^{d0'}_{g}$ is the time constant related to the flux decay of the machine. Finally, the active and reactive power output of DG $g$ are expressed as
\begin{equation}
P^{e}_{g} = v^d_{g} i^d_{g} + v^q_{g} i^q_{g},\forall g \in \mathcal{DG},
\end{equation}
\begin{equation}
Q^{e}_{g} = v^q_{g} i^d_{g} - v^d_{g} i^q_{g},\forall g \in \mathcal{DG}.
\end{equation}

\subsubsection{Inverter-based Resources}
We assume that IBRs are operated in PQ mode, synchronised with the HMG’s frequency \cite{jiang2023dynamics}. The voltage and current control loops of IBRs are decoupled, while DC/DC converters are modelled as an ideal pure power injection. Thus, the $q$ axis of IBRs is aligned with the terminal voltage, i.e., $v^q_g \!=\! V_b$ and $v^d_g \!=\! 0$. The current references for $d$ and $q$ axes can be calculated from active and reactive power references $P^{ref}_g$ and $Q^{ref}_g$, expressed as
\begin{equation}
i^{q,ref}_{g} = P^{ref}_{g}/v^q_g,\forall g \in \mathcal{IBR},
\end{equation}
\begin{equation}
i^{d,ref}_{g} = Q^{ref}_{g}/v^q_g,\forall g \in \mathcal{IBR}.
\end{equation}

To control both $d$ and $q$ axes independently, a Proportional-Integral (PI) current controller is adopted with a feed-forward term for input disturbance rejection \cite{kroutikova2007state}. In detail, the control scheme can be simplified as the following two differential equations expressed in the Laplace domain:
\begin{equation}
si^{d}_{g} = 1/L(K_{P}+K_{I}/s)(i^{d,ref}_{g}-i^{d}_{g}),\forall g \in \mathcal{IBR},
\end{equation}
\begin{equation}
si^{q}_{g} = 1/L(K_{P}+K_{I}/s)(i^{q,ref}_{g}-i^{q}_{g}),\forall g \in \mathcal{IBR},
\end{equation}
where $L$ is the smoothing output inductance, and $K_P$ is the proportional controller gain. Finally, similar to DGs, the active and reactive power outputs of IBR $g$ can be expressed as
\begin{equation}\label{eq:in_p}
P^{ibr}_{g} = v^q_{g} i^q_{g},~Q^{ibr}_{g} = v^q_{g} i^d_{g},\forall g \in \mathcal{IBR}.
\end{equation}
\vspace{-1.48em}

\subsection{Controller Dynamic Constraints}
\label{sec:III.B}
To mitigate post-fault frequency deviation and rectify the instantaneous power imbalance, frequency control loops based on PI control are introduced into the proposed TSC-OPF for controllable DGs, as depicted in Fig. \ref{fig:cd}. In addition, voltage control is achieved via the automatic voltage regulator (AVR) of DGs and the voltage droop controller of IBRs. Details regarding the introduced frequency and voltage control strategies are presented below:

\begin{figure}[t!] 
\vspace{-0.18em}
\centering  
\includegraphics[width=0.49\textwidth]{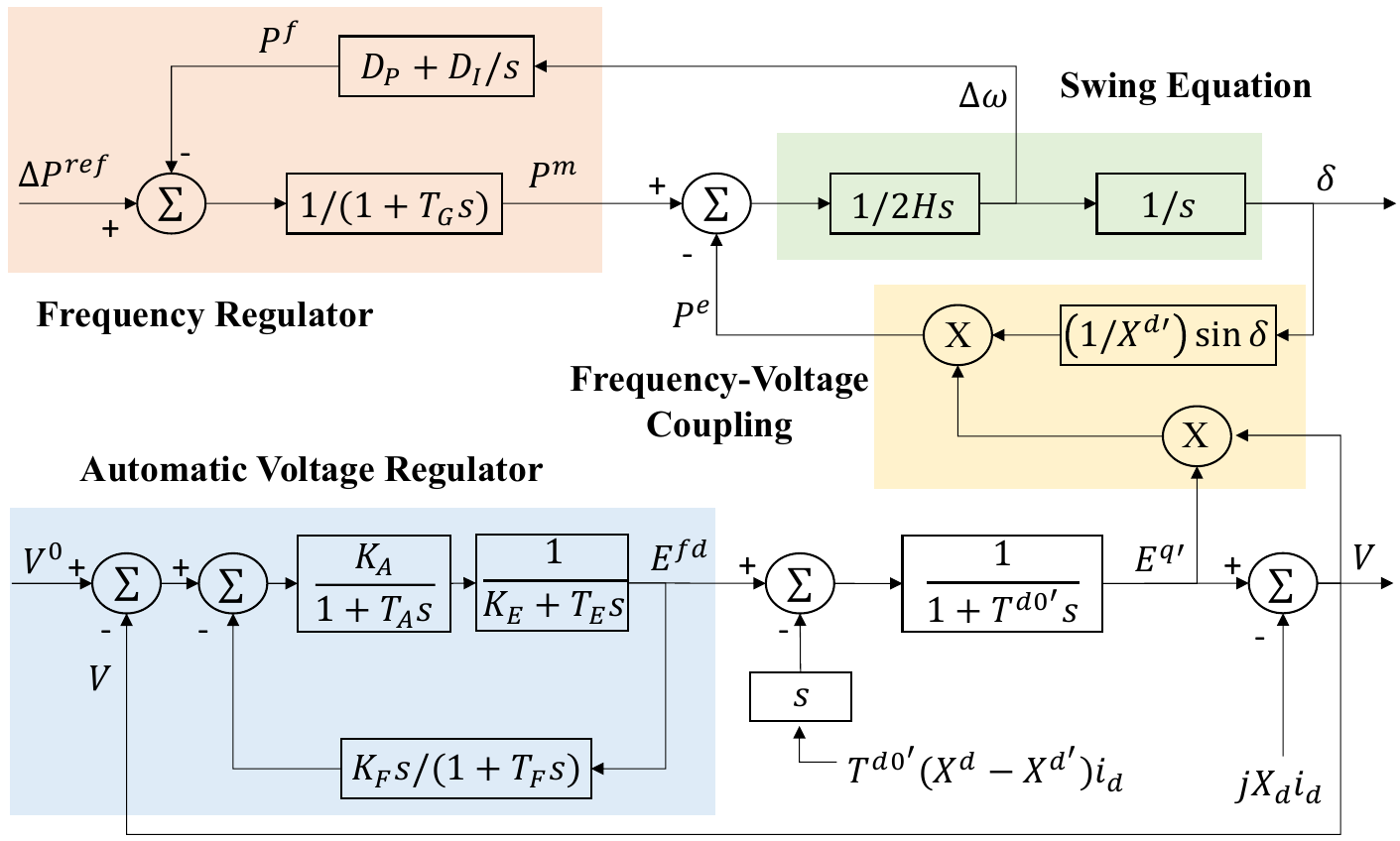}
\vspace{-1.28em}
\caption{Frequency and voltage control loops integrated with the TSC-OPF for conventional generators.}
\label{fig:cd}
\vspace{-1.28em}
\end{figure}

\subsubsection{Frequency Control}
As shown in Fig. \ref{fig:cd}, a PI control loop is introduced to dispatch DG and ensure a steady-state frequency (e.g., 50 Hz). The PI control action for DG $g$ in the Laplace domain can be expressed as
\begin{equation}\label{eq:fre_con_1}
P^{f}_{g} = -(D_{P} + D_{I}/s) \Delta \omega_{g},\forall g \in \mathcal{DG},
\end{equation}
where $P^{f}_{g}$ corresponds to the outputted power set-point actuating the throttle valve of the prime mover. $D_{P}$ and $D_{I}$ are the proportional and integral parameters of the PI controller, respectively. Furthermore, the dynamics of the prime mover can be approximated by the following differential equation:
\begin{equation}\label{eq:fre_con_2}
P^{m}_{g} =  P^{f}_{g}/(1+T_{G} s),\forall g \in \mathcal{DG},
\end{equation}
where $T_{G}$ refers to the charging time constant.

\subsubsection{Voltage Control}
AVRs are designed for the primary voltage regulation of controllable DGs, ensuring that the bus voltage is close to the nominal value (e.g., 1 $p.u.$). In this section, AVR Type I with a standard DC exciter IEEE Type I \cite{ieee} is adopted for voltage regulation, while an additional stabilising feedback block is included. The error between the measured voltage and the reference voltage is fed into an amplifier and an exciter sequentially, followed by a feedback loop based on a first-order stabiliser model, as specified in the IEEE standard 421.5 \cite{ieee}.

In detail, the exciter and stabiliser closed-loop transfer function can be expressed in the Laplace domain as
\begin{equation}
\begin{split}
  &E^{r}_{g}(T_F s +1)+T_{A}sE^{r}_{g}(T_F s +1) = \\ 
  &K_A(V^{0}_{b}-V_{b}) -T_F K_A s V_{b} - K_F K_A s E^{fd}_{g},  
\end{split}
\end{equation}
\begin{equation}
  E^{fd}_{g} = E^{r}_{g}/(K_{E}+T_{E}s),\forall g \in \mathcal{DG}, \forall b \in \mathcal{B}_{dg},
\end{equation}
where $K_A$ and $T_A$ are the amplifier gain and time constant. $K_E$ and $T_E$ are the field circuit integral deviation and time constant, while $K_F$ and $T_F$ refer to the stabiliser gain and time constant, respectively. $E^{r}_{g}$ is the regulator voltage. Furthermore, the reactive power outputs of IBRs can be controlled by a proportional voltage droop controller, where the reactive power outputs are defined as a function of voltage magnitude:
\begin{equation}\label{eq:in_con}
Q^{ibr}_{g} = Q^{ibr,0}_{g}+D_{V}(V^{0}_{b}-V_{b}),\forall g \in \mathcal{IBR}, \forall b \in \mathcal{B}_{ibr},
\end{equation}
where $Q^{ibr,0}_{g,t}$ is the nominal reactive power of IBRs. $D_{V}$ denotes the droop constant of IBRs \cite{aprilia2019unified}.

\subsection{Hybrid Power Flow of the HMG}
\vspace{-0.0em}
\label{sec:III.C}
\subsubsection{Power Exchange between AC and DC Subgrids}
\vspace{-0.0em}
\label{sec:III.C}
In the HMG system, AC and DC subgrids can exchange active power through AC/DC ICs. On the AC side, the active power is associated with frequency, while the active power on the DC side depends on voltage. Owing to their different scales, AC frequency and DC bus voltage can only be compared using the normalised values \cite{aprilia2019unified}. Therefore, a normalisation technique known as `feature scaling' in statistics is employed to normalise AC frequency and DC bus voltage as
\begin{equation}\label{eq:exchange_1}
\hat \omega_{i} = \frac{2\cdot \omega_{i} - (\overline{\omega}_{i}+\underline{\omega}_{i})}{\overline{\omega}_{i}-\underline{\omega}_{i}}, \forall i \in \mathcal{MG}^{ac},
\end{equation}
\begin{equation}\label{eq:exchange_2}
\hat V^{dc}_{j} = \frac{2\cdot V^{dc}_{j} - (\overline{V}^{dc}_{j}+\underline{V}^{dc}_{j})}{\overline{V}^{dc}_{j}-\underline{V}^{dc}_{j}},\forall j \in \mathcal{MG}^{dc},
\end{equation}
where the true values of AC frequency and DC voltage are normalised to a range of $[-1,1]$, facilitating the comparison of $\hat \omega_{i}$ and $\hat V^{dc}_{j}$. This normalisation process effectively correlates AC frequency with DC voltage, enabling the subsequent power exchange between AC and DC subgrids.

\begin{figure}[t!] 
\vspace{-0.18em}
\centering  
\includegraphics[width=0.45\textwidth]{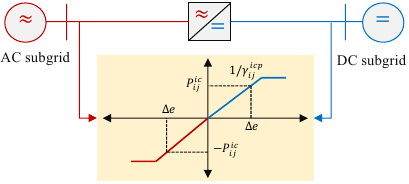}
\vspace{-0.48em}
\caption{Droop control strategy of the AC/DC interlinking converter.}
\label{fig:droop}
\vspace{-1.3em}
\end{figure}

Consequently, following the droop control scheme demonstrated in Fig. \ref{fig:droop}, the power exchange $P^{ic}_{ij}$ through the IC connecting AC subgrid $i$ and DC subgrid $j$ is defined as
\begin{equation}
P^{ic}_{ij} = -\frac{1}{\gamma^{icp}_{ij}}(\hat \omega_{i} - \hat V^{dc}_{j}),\forall i \in \mathcal{MG}^{ac},\forall j \in \mathcal{MG}^{dc},
\end{equation}
where $\gamma^{icp}_{ij}$ is the droop constant for frequency-voltage gain. When active power flows from the DC side to the AC side, the converter can inject reactive power to the AC subgrid. This injected reactive power can be regulated using the following droop control scheme related to the IC capacity:
\begin{equation}\label{eq:ic_0}
Q^{ic}_{ij} \leq \sqrt{(\overline{S}^{ic}_{ij})^{2}-(P^{ic}_{ij})^2}, \forall i \in \mathcal{MG}^{ac},\forall j \in \mathcal{MG}^{dc},
\end{equation}
\begin{equation} \label{eq:ic_1}
    Q^{ic}_{ij}=
    \begin{cases}
    Q^{ic,0}_{i}+\frac{1}{\gamma^{icq}_{ij}}(V^{ac,0}_{i}-V^{ac}_{i}),& \text{$P^{ic}_{ij}>0$} \\
    0, & \text{$P^{ic}_{ij} \leq 0$}
    \end{cases},
\end{equation}
where $\gamma^{icq}_{ij}$ is the droop constant of IC reactive power-AC voltage gain. $Q^{ic,0}_{i}$ corresponds to nominal IC reactive power.

\subsubsection{Network Operation}
\vspace{-0.0em}
\label{sec:III.C}
The operation of the HMG network can be fully modelled using the following AC/DC hybrid power flow constraints, including both AC and DC sides. For AC subgrids, active and reactive power injections at AC bus $b$ in set $\mathcal{B}^{ac}$ can be expressed as
\begin{equation} \label{eq:active} 
    P^{ac}_{b} = V^{ac}_{b}\sum_{(b,p) \in \mathcal{L}^{ac}}V^{ac}_{p}Y_{bp}\cos(\delta_{b}-\delta_{p}-\delta_{bp}),
\end{equation}
\begin{equation} \label{eq:reactive} 
    Q^{ac}_{b} = V^{ac}_{b}\sum_{(b,p) \in \mathcal{L}^{ac}}V^{ac}_{p,t}Y_{bp}\sin (\delta_{b}-\delta_{p}-\delta_{bp}),
\end{equation}
where $Y_{bp}$ is admittance matrix. Unlike a system with a stiff grid, $Y_{bp}$ is not constant, expressed as a function of frequency:
\begin{equation} \label{eq:admittance} 
    Y_{bp}(\omega)= Y_{bp}(\omega)e^{j\theta_{bp}(\omega)}=-Z^{-1}_{bp}(\omega),
\end{equation}
where $Z^{-1}_{bp}(\omega)=R_{bp}+jX_{bp}(\omega)$ is the line impedance between AC buses $b$ and $p$. Similarly, in DC subgrids, the power injection of DC bus $b$ in set $\mathcal{B}^{dc}$ is constrained by
\begin{equation} \label{eq:dc_active} 
    P^{dc}_{b} = V^{dc}_{b}\sum_{(b,p) \in \mathcal{L}^{dc}}V^{dc}_{p}G_{bp},
\end{equation}
where $G_{bp}$ refers to the line conductance between DC buses $b$ and $p$. $\mathcal{L}^{ac}$ and $\mathcal{L}^{dc}$ correspond to the line sets of AC and DC subgrids in the HMG, respectively.
\vspace{-0.8em}

\subsection{Static Generation Constraints of the HMG}
\label{sec:III.A}
Except for dynamic constraints, the proposed TSC-OPF considers the following static constraints for time horizon $T$:

\subsubsection{Conventional Generator}
\label{sec:III.D}
The scheduling behaviours of DGs on AC side are modelled as
\begin{equation}\label{eq:gen energy range}
    \underline{P}_{g}^{dg} \leq P_{g,t}^{dg} \leq \overline{P}_{g}^{dg},\forall g \in \mathcal{DG}^{ac},\forall t \in T,
\end{equation}
\begin{equation}\label{eq:gen apparent}
    (P^{dg}_{g,t})^2+(Q^{dg}_{g,t})^2 \leq (\overline{S}^{dg}_{g})^{2}, \forall g \in \mathcal{DG}^{ac}, \forall t \in T,
\end{equation}
\begin{equation}\label{eq:gen pf}
    |Q^{dg}_{g,t}| \leq P^{dg}_{g,t} \tan(\cos^{-1}\delta^{dg}_{g}),\forall g \in \mathcal{DG}^{ac},\forall t \in T,
    \vspace{-0.03cm}
\end{equation}
where \eqref{eq:gen energy range} and \eqref{eq:gen apparent} define the active and reactive power output limits of DG $g$ influenced by the rated power factor $\delta^{dg}_{g}$. Note that the above constraints are used for the AC subgrid, while only active power constraints are applied to the DC subgrid.

\subsubsection{Renewable Energy Resource}
\label{sec:III.A.3}
Similarly, the operating model of IBRs (e.g., WTs and PVs) can be expressed as
\begin{equation}\label{acpowerres}
    0 \leq P^{ibr}_{g,t} \leq \alpha_{g,t}\overline{P}^{ibr}_{g}, \forall g \in \mathcal{IBR}^{ac}, \forall t \in T,
\end{equation}
\begin{equation}\label{acpowerres_2}
    (P^{ibr}_{g,t})^2+(Q^{ibr}_{g,t})^2 \leq (\overline{S}^{ibr}_{g})^{2}, \forall g \in \mathcal{IBR}^{ac}, \forall t \in T,
\end{equation}
where constraint \eqref{acpowerres} limits the active power of IBR $g$, affected by weather conditions (e.g., wind speed) contained in $\alpha_{g,t}$.

\subsubsection{Energy Storage}
\label{sec:III.A.2}
ESs in the HMG are modelled as
\begin{equation}\label{eq:ev model charge}
    0 \leq P^{c}_{k,t} \leq u^{es}_{k,t} \overline{P}^{es}_{k}, \forall k \in \mathcal{ES}, \forall t \in T,
\end{equation}
        \vspace{-0.2cm}
\begin{equation}\label{eq:ev model discharge}
    (u^{es}_{k,t}-1) \overline{P}^{es}_{k} \leq P^{d}_{k,t} \leq 0, \forall k \in \mathcal{ES}, \forall t \in T,
\end{equation}
        \vspace{-0.2cm}
\begin{equation}\label{eq:ev reactive}
    (P^{d}_{k,t}+P^{c}_{k,t})^2 + (Q^{es}_{k,t})^2 \leq (\overline{S}^{es}_{k})^2, \forall k \in \mathcal{ES}, \forall t \in T,
\end{equation}
        \vspace{-0.2cm}
\begin{equation}\label{eq:ev storage energy}
    \underline{E}^{es}_{k} \leq E_{k,t}^{es} \leq \overline{E}^{es}_{k}, \forall k \in \mathcal{ES}, \forall t \in T,
\end{equation}
        \vspace{-0.2cm}
\begin{equation}\label{eq:ev storage soc}
\begin{split}
\!\!\!\!\!\!\!E_{k,t+1}^{es} \!=\! E_{k,t}^{es} + (P^{c}_{k,t}\eta_{k}^{es} \!+\! P^{d}_{k,t}/\eta_{k}^{es}) \Delta t, \\ \forall k \in \mathcal{ES}, \forall t \in T,
\end{split}
\end{equation}
where constraints \eqref{eq:ev model charge}-\eqref{eq:ev storage soc} restrict charging/discharging behaviours and energy content of ES $k$. $u^{es}_{k,t}$ refers to the status of ES $k$ at time step $t$ ($u^{es}_{k,t}=1$ if charging; 0, otherwise).
\vspace{-0.2cm}

\subsection{From Operation to Planning}
\label{sec:III.D}
\subsubsection{Objective Function}
Since we are dealing with a planning problem, the main objective function should be to minimise overall costs, including both annuitized infrastructure investment costs and annual system operation costs, where the investment costs can be defined as
\begin{equation} \label{eq:obj}
\begin{split}
F^{in}=&\sum_{g \in \mathcal{DG}} G^{dg}_{g} (C^{gen}_{g}AF^{gen}_{g}+C^{fix,gen}_{g})\\
&+\sum_{g \in \mathcal{IBR}} G^{ibr}_{g} (C^{ibr}_{g}AF^{ibr}_{g}+C^{fix,ibr}_{g})\\
&+\sum_{k \in \mathcal{ES}} G^{es}_{k} (C^{es}_{k}AF^{es}_{k}+C^{fix,es}_{k}),
\end{split}
\end{equation}
including capital costs of DGs, IBRs (WTs and PVs), and ESs. Annuity factors (e.g., $AF^{gen}_{g}$) are considered for different assets according to their discount rates and life spans, while $C^{fix,gen}_{g}$ corresponds to the fixed operating and maintenance cost \cite{zhang2018whole}. The operation costs $F^{op}$ of the HMG are defined as
\begin{equation}\label{eq:op_obj}
\begin{split}
    \!\!F^{op} = \sum_{g \in \mathcal{DG}^{ac}}\sum_{t \in T} c^{gen}_{g} P_{g,t}^{dg,ac}
    +\sum_{g \in \mathcal{DG}^{dc}}\sum_{t \in T} c^{gen}_{g} P_{g,t}^{dg,dc},
\end{split}
\end{equation}
where $c^{gen}_{g}$ corresponds to the generation cost of DG $g$. Furthermore, to ensure the sizing results can maintain transient stability, the following constraints should be considered \cite{zhao2020frequency}:
\begin{equation} \label{eq:tsc_con}
\begin{cases}
    \underline{f} \leq f^{s} \leq \overline{f},\\
    \underline{V} \leq V \leq \overline{V},
\end{cases}
\end{equation}
where $\underline{f},\overline{f}$ are the allowed minimum/maximum system frequency. $\underline{V},\overline{V}$ correspond to voltage limits.

Within the DAD architecture, the key objective for the attacker should be to find the worst-case scenario for maximising frequency and voltage deviations, which can be expressed as
\begin{equation} \label{eq:tsc_obj}
F^{atk}= \Delta f^{s} + \alpha\Delta V,
\end{equation}
where $\Delta f^{s}$ and $\Delta V$ refer to frequency and voltage deviations during the contingency. $\alpha$ corresponds to a weighting factor. The primary purpose of introducing the weighting factor is to harmonise the scales of frequency and voltage deviations within the objective function \eqref{eq:tsc_obj}, ensuring that both are evaluated on a comparable basis. In detail, the weighting factor $\alpha$ is chosen to balance the relative magnitudes of frequency and voltage deviations, since they may be expressed in different units, e.g., Hz for frequency and p.u. for voltage. After introducing the weighting factor, the objective function can integrate these deviations by normalising their values according to typical ranges or standard deviations observed in the system. This normalisation ensures that changes in frequency and voltage are considered equitably, allowing for a holistic representation of the system's stability. This approach not only enhances the precision of the proposed attacker model with objective function \eqref{eq:tsc_obj} but also eventually ensures that the planning decisions are stable, resilient and adaptive to varying system conditions, particularly when heuristic algorithms such as GA are employed to explore a range of potential scenarios.

According to \cite{hatziargyriou2020definition}, frequency stability refers to the ability of a power system to maintain a consistent frequency, especially in the presence of fluctuations or disturbances. Voltage stability refers to the ability of a power system to maintain steady voltages at all buses in the system after being subjected to a disturbance from a given initial operating condition. Therefore, the main focus of the above definitions is to maintain frequency and voltage within acceptable ranges (e.g., around 50 Hz for frequency and around 1 p.u. for voltage). In this context, the proposed TSC-OPF model incorporates both the advantages of mathematical models and time-domain simulations, which is capable of accurately capturing second-scale frequency and voltage deviations during various contingencies; thus, monitoring frequency and voltage deviations for each second and realising them as indicators for stability verification can be a straightforward and reliable method.

\begin{figure}
\vspace{-0.15cm}
\centering
\includegraphics[width=0.3\textwidth]{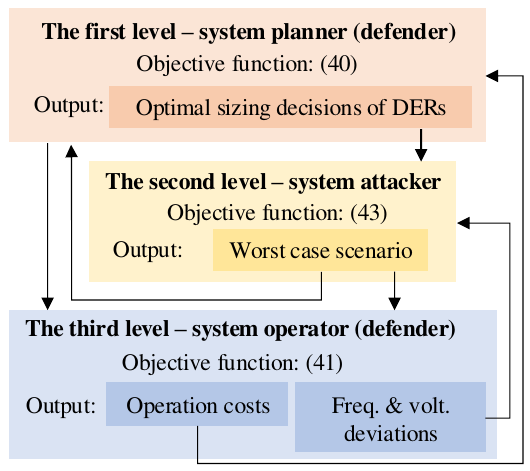}
\vspace{-0.2cm}
\caption{The initial formulation of the proposed three-level DAD model and corresponding mutual interactions.}\label{fig:three_level}
\vspace{-0.4cm}
\end{figure}

\subsubsection{Final Mathematical Formulation}
Based on the above objectives, the DAD model can be divided into three levels, i.e., the first-level planner (defender), the second-level attacker, and the third-level operator (defender), where detailed information of each level and their mutual interactions are illustrated in Fig. \ref{fig:three_level}. This three-level setting including both planner and operator allows the HMG system to have the ability to make different reactions when different contingencies happen, which better simulates real-world practices. Afterwards, the DAD model can be effectively reformulated or decomposed into an upper-level problem and a low-level problem \cite{wang2021three}, expressed as follows:

a) \textbf{Upper-level Problem}: the defender model is written as:
\begin{equation}\label{eq:upper}
\begin{split}
obj.~
 \min_{G^{dg},G^{ibr},G^{es}}~\eqref{eq:obj}~ \min_{P,Q}~\eqref{eq:op_obj}~~~~~~~~~~~~\\
    ~~s.t.~\begin{cases}\textit{dynamic generator constraints:}~\eqref{eq:rotor}-\eqref{eq:in_p},\\
    \textit{controllor dynamic constraints}~\eqref{eq:fre_con_1}-\eqref{eq:in_con},\\
    \textit{hybrid power flow constraints:}~\eqref{eq:exchange_1}-\eqref{eq:dc_active},\\
    \textit{static generation constraints:}~\eqref{eq:gen energy range}-\eqref{eq:ev storage soc},\\
    \textit{transient stability constraint:}~\eqref{eq:tsc_con},
    \end{cases}
\end{split}
\end{equation}

b) \textbf{Low-level Problem}: the attacker model is written as:
\begin{equation}\label{eq:lower}
\begin{split}
 obj.~&\max_{AD}~\eqref{eq:tsc_obj}~ \min_{P,Q}~\eqref{eq:op_obj}\\
    &~s.t.~\eqref{eq:rotor}-\eqref{eq:ev storage soc},
\end{split}
\end{equation}
where $AD$ refers to a set of attack actions, such as wind power shortage, IC outage, etc. Note that the low-level problem is formulated in a $max\!-\!min$ fashion, aiming to find the worst case, e.g., the largest frequency and voltage deviation. Overall, as depicted in Fig. \ref{fig:ccg}, the upper-level and low-level problems will be run iteratively until convergence (e.g., meeting stability requirements under any simulated contingency scenarios), while a detailed solving procedure is provided in Section \ref{sec:IV}. It is worth noting that the planning decisions (e.g., $G^{dg},G^{ibr},G^{es}$) for the low-level problem \eqref{eq:lower} should be the same as the final sizing decisions of the upper-level problem \eqref{eq:upper} within the same iteration, since this paper primarily focuses on the planning perspective and aims to analyse the effectiveness of the optimal sizing decisions under different contingencies.
\vspace{-0.2cm}

\section{Solving Procedure}
\label{sec:IV}
The proposed DAD model can be solved via a column-and-constraint generation (C\&CG) method until convergence, where $\epsilon$ corresponds to the value of convergence tolerance \cite{wang2021three,zhang2020multi}. However, to effectively solve the above planning problem, three challenges need to be appropriately addressed:

\begin{figure}
\vspace{-0.2cm}
\centering
\includegraphics[width=0.45\textwidth]{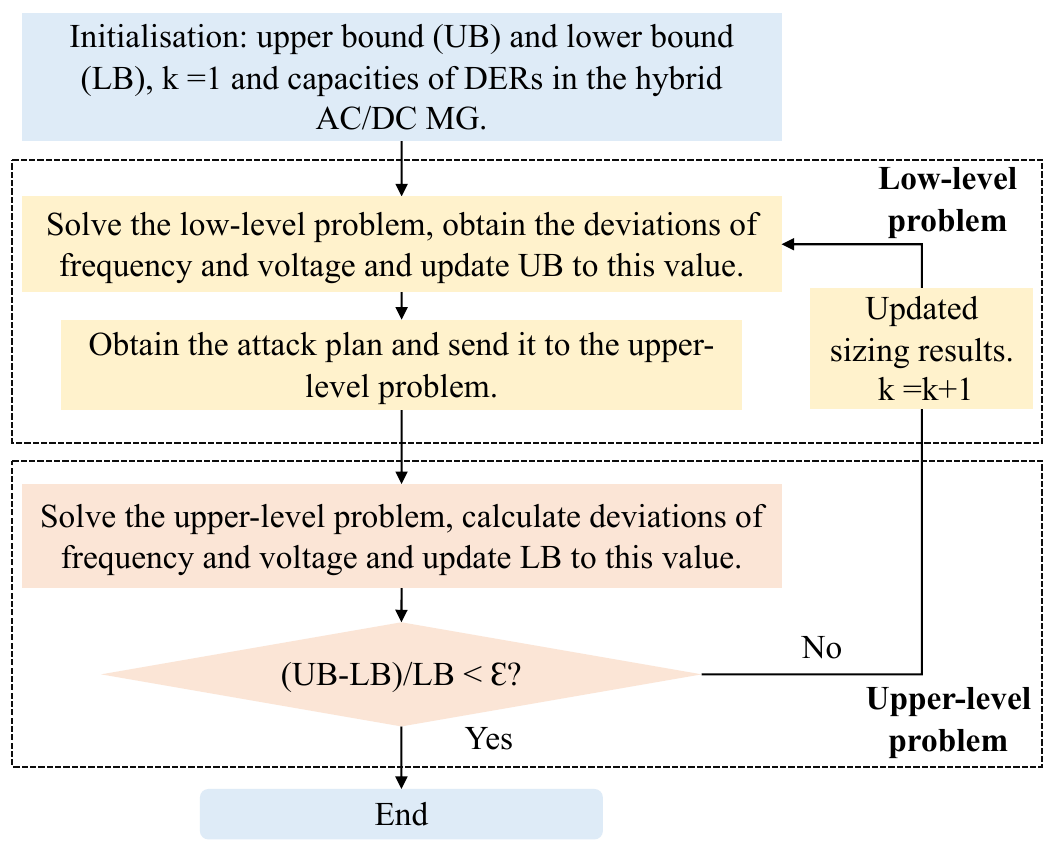}
\vspace{-0.1cm}
\caption{Solving procedure of the proposed DAD planning framework.}\label{fig:ccg}
\vspace{-0.1cm}
\end{figure}

1) TSC-OPF is a DAE-constrained optimisation problem including both algebraic and differential equations, which cannot be directly co-optimised by optimisation algorithms. In detail, the transient nature of differential equations necessitates a second-scale resolution across various system conditions, while traditional optimisation software such as IPOPT and Gruobi typically address static or quasi-static states and cannot directly handle the dynamic evolution of system states.

2) Static constraints and differential equations deal with different time domains. The time-coupled static model is used to capture the gradual change of RES and load profiles (e.g., hourly changes) and then make economic decisions, while the dynamic model is used to capture the sudden change of RES or load on a second scale to simulate frequency and voltage deviations. Thus, directly linking the daily optimisation problem with second-level dynamics can be difficult.

3) Considering optimal sizing decisions, non-linear power flow, and differential equations makes the optimisation problem become mixed-integer non-linear programming (MINLP), which cannot be efficiently solved by commercial software.

Therefore, it can be found from Fig. \ref{fig:scale} that there exist three different time scales in the proposed stability-driven optimal sizing model, including planning decisions (yearly time scale with daily resolution), time-coupled daily decisions (daily time scale with hourly resolution), and TSC-OPF with frequency and voltage stability (hourly time scale with second resolution). To address the above challenges and consider these different time scales in one unified planning framework, the following corresponding solving strategies are proposed:
\vspace{-0.4cm}

\begin{figure}
\vspace{-0.1cm}
\centering
\includegraphics[width=0.46\textwidth]{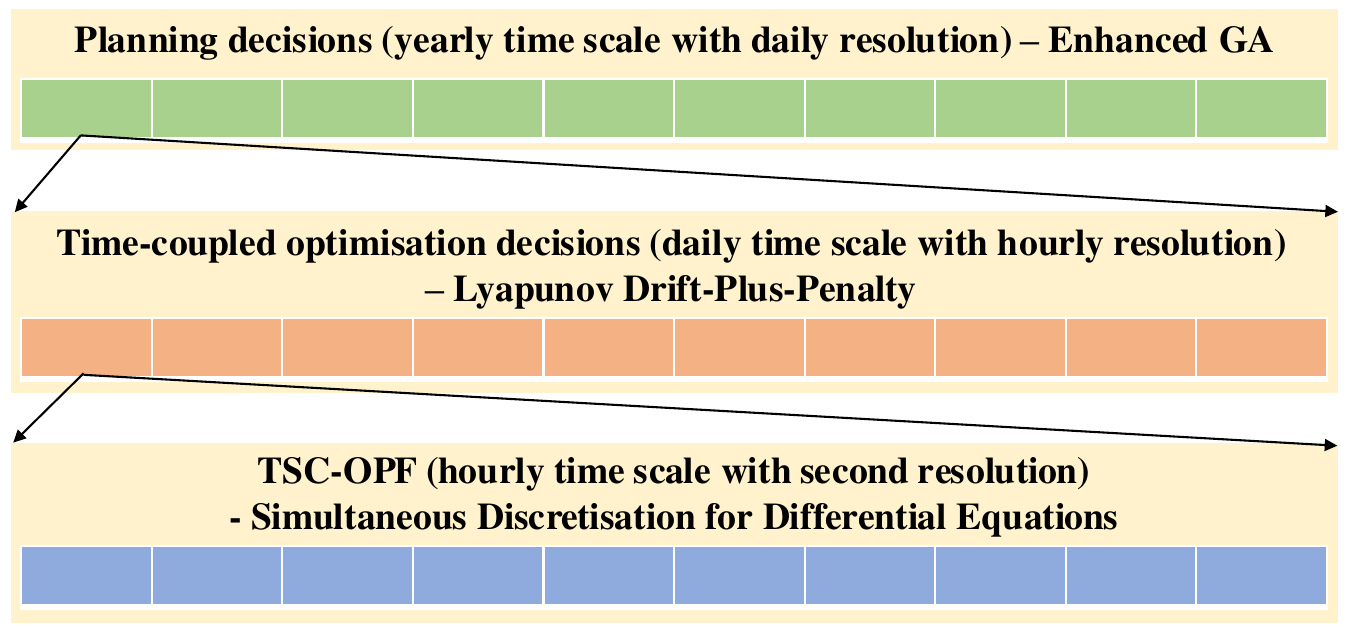}
\vspace{-0.1cm}
\caption{Three different time scales in the proposed stability-driven optimal sizing model.}\label{fig:scale}
\vspace{-0.3cm}
\end{figure}

\subsection{Simultaneous Discretisation for Differential Equations}
\label{sec:IV.A}
To solve differential equations, a simultaneous discretisation technique is utilised, where the key is to use algebraic equations to approximate differential equations. In detail, this method can approximate the differential equation at a particular point via a difference equation. Depending on the points used for approximation, there are various types \cite{zhao2020frequency}, where the backward difference, also known as backward Euler, is a common one that works as below:
\begin{equation}\label{eq:encoder}
\begin{split}
    (\frac{dx_t}{dt},f(x_t,u_t))=0, t \in [0,T],\\
    \frac{dx_t}{dt}|_{t_{k+1}}=\frac{x_{k+1}-x_{k}}{h}=0, k=0,...,N-1,\\
    g(\frac{dx_t}{dt}|_{t_{k+1}},f(x_{k+1},u_{k+1}))=0, k=0,...,N-1,\\
\end{split}
\end{equation}
where $x_{k}=x_{t_k}$ and $t_{k}=kh$. $h$ is the step size between discretisation points. In this case, differential equations can be transferred into algebraic equations for co-optimisation.
\vspace{-0.2cm}

\subsection{Lyapunov Drift-Plus-Penalty for Time-Coupled Operation}
\label{sec:IV.B}
To capture sudden changes and transient stability, it is necessary to solve the TSC-OPF for every static period (e.g., each hour) in a time-decoupled manner. Then, the time-coupling features (e.g., ESs) in a daily optimisation can be handled by Lyapunov Drift-Plus-Penalty optimisation \cite{shi2022lyapunov}. In general, Lyapunov optimisation was first introduced to optimise the management of queueing systems, aiming to strike a balance between system performance and queue congestion. Regarding the proposed time-coupled optimisation problem, the energy contents of ESs can be analogous to queue backlogs. The goal of Lyapunov optimisation is to maintain a good balance between ES energy in \eqref{eq:ev storage soc} and operation costs in \eqref{eq:op_obj}.

To fit the problem into a Lyapunov optimisation framework, we first define a queue $Q_{k,t}$ for ES $k$ as
\begin{equation}\label{eq:encoder}
    Q_{k,t} = \overline{E}^{es}_{k}-E^{es}_{k,t}, \forall t \in T, 
\end{equation}
where the queue $Q_{k,t}$ indicates the amount of discharged energy of ES $k$. To comply with the energy capacity of ES $k$ in \eqref{eq:ev storage energy}, $Q_{k,t}$ should be restricted within $[0,\overline{E}^{es}_{k}-\underline{E}^{es}_{k}]$.

Based on the time-coupling feature of ES energy and the queue definition, $Q_{k,t}$ can be updated as
\begin{equation}\label{eq:encoder}
    Q_{k,t+1} = Q_{k,t}+U_{k,t} = Q_{k,t}+(P^{d}_{k,t}-P^{c}_{k,t})\Delta T, \forall t \in T,
\end{equation}
where $U_{k,t}$ denotes the input of the queue at time step $t$. Then, a quadratic Lyapunov function is defined for all ESs as
\begin{equation}\label{eq:encoder}
    L(\Theta_{t})= 1/2 \sum_{k \in \mathcal{ES}} (Q_{k,t})^2, \forall t \in T,
\end{equation}
where $\Theta_{t}\! =\! [Q_{1,t},...,Q_{|\mathcal{ES}|,t}]$. To capture increment of queues' backlog, one-slot conditional Lyapunov drift is derived as
\begin{equation}\label{eq:encoder}
    \Delta \Theta_{t}=\mathrm{E}[L(\Theta_{t+1})-L(\Theta_{t})|\Theta_{t}], \forall t \in T,
\end{equation}
where the Lyapunov drift describes the expected change of $L(\Theta_{t})$ for one-time step, indicating ES energy change.

To this end, using Lyapunov drift $L(\Theta_{t})$, original operation problem \eqref{eq:op_obj} for each time step $t$ is reformulated as
\begin{equation}\label{eq:op_obj_middle}
\begin{split}
    \min \Delta J_t = &\mathbb{E}[\Delta \Theta_{t}+V \cdot F^{op}_{t}|\Theta_{t}]\\
    &s.t.~\eqref{eq:rotor}-\eqref{eq:ev storage soc},
\end{split}
\end{equation}
which refers to drift-plus-penalty as a weighted sum of the Lyapunov drift and the expected immediate cost, given the current queue backlog. $V$ is a weighting parameter.

Due to Lyapunov drift, the optimisation \eqref{eq:op_obj_middle} is intractable and cannot be directly solved. Therefore, instead of the drift-plus penalty, an upper bound linked with constant $B$ is introduced and then minimised, which can be defined below: 

\textbf{Lemma 1}: The one-slot conditional Lyapunov drift $L(\Theta_{t})$ satisfies the following inequality for each time step $t$:
\begin{equation}\label{eq:encoder}
    \Delta \Theta_{t} \leq B + \sum_{k \in \mathcal{ES}}Q_{k,t}\mathbb{E}(U_{k,t}|Q_{k,t}).
\end{equation} 

Following \textbf{Lemma 1}, the objective in \eqref{eq:op_obj_middle} can be reduced to a deterministic expression for an arbitrary time step $t$ as
\begin{equation}\label{eq:op_obj_2}
\begin{split}
\Delta J_t \leq B + \sum_{k \in \mathcal{ES}}Q_{k,t}U_{k,t}+V \cdot F^{op}_{t}, \forall t \in T.
\end{split}
\end{equation}

Finally, the objective in \eqref{eq:op_obj_middle} can be converted by minimising the following upper bound:
\begin{equation}\label{eq:op_obj_new}
\begin{split}
 \min \sum_{k \in \mathcal{ES}} & Q_{k,t}U_{k,t}+V \cdot F^{op}_{t},
\end{split}
\end{equation}
where the queue backlog $Q_{k,t}$ corresponds to the discharged battery energy. The weighting parameter $V$ is used for the trade-off between queue backlog reduction and immediate cost \cite{shi2022lyapunov}. By adjusting $V$, the system can effectively choose to prioritise immediate operation costs or ES energy content, eventually achieving a balance between these two aspects and long-term cost-effective solutions. In detail, a larger value of $V$ represents minimising the immediate operation costs at the cost of prolonging the queue length, while a smaller value of $V$ represents minimising the queue length at the cost of slightly higher operation costs \cite{tian2024lyapunov}. Specifically, multiple $V$s can be incorporated for different periods, such as peak and non-peak demand hours. 
\vspace{-0.3cm}

\subsection{Enhanced GA with Sparsity Calculation for MINLPs}
\label{sec:IV.C}
To deal with the MINLP problems in both the upper-level ($min-min$) \eqref{eq:upper} and low-level ($max-min$) \eqref{eq:lower}, an adaptive genetic algorithm (AGA) with sparsity calculation and local search is developed to further divide the upper-level and low-level problems into two stages by handling integer variables relating to DER capacities (upper-level \eqref{eq:upper}) or potential contingencies (low-level \eqref{eq:lower}). For example, the enhanced GA can be used to generate contingency scenarios and simulate the `$max$’ function in \eqref{eq:lower}. The generated scenarios will be sent to the HMG operator with the `$min$’ function in \eqref{eq:lower} to minimise operation costs under different contingencies, which is solved by the proposed TSC-OPF algorithm. In this way, low-level problem \eqref{eq:lower} can be effectively solved by the combination of enhanced GA and TSC-OPF model. The process of the enhanced GA is depicted in Fig. \ref{fig:aga}, while algorithm details are given below:

\subsubsection{Adaptive Genetic Algorithm}
GA is categorised as a global search meta-heuristic that does not suffer from parameter numbers and is efficient in terms of computational time and programming simplicity \cite{rousis2019planning}. However, traditional GA, which normally adopts a small mutation rate (e.g., 0.01-0.1), may easily fall within the local minimum, while a large mutation rate (e.g., over 0.3) can make the problem hard to converge. In this context, this paper introduces AGA with adaptive crossover and mutation probabilities to avoid local optimum and speed up convergence. In this case, populations with different fitness values can have different probabilities for GA update. The crossover and mutation probabilities ($p^c$ and $p^m$) are calculated as
\begin{equation} \label{adaptive}
\begin{cases}
p^c = k_1 \frac{F^{max} - F^{ave}}{F^{max}-F'},~when~F' \leq F^{ave},\\
p^c = k_3,~when~F' > F^{ave},\\
p^m = k_2 \frac{F^{max} - F^{ave}}{F^{max}-F'},~when~F' \leq F^{ave},\\
p^m = k_4,~when~F' > F^{ave},
\end{cases}
\end{equation}
where $F'$, $F^{ave}$, and $F^{max}$ correspond to the smaller fitness value in the operation of crossover, the average fitness value, and the largest fitness value within the population, respectively. $k_1$, $k_2$, $k_3$, $k_4$ are constants, where the choice of their values is empirical and can be changed in different problems \cite{wang2021three}.

\subsubsection{Sparsity Calculation and Local Search}
To enhance the diversity of the GA population, sparseness theory and local search \cite{luan2023enhanced} are incorporated into the AGA, which allows the algorithm to obtain new children populations by performing local search operations on the sparse solution. Thus, both population diversity and algorithm generality can be improved.

In detail, we first calculate the Euclidean distance between one individual and other individuals, expressed as
\begin{equation}
    \rho = \sqrt{\sum_{i=1}^{N}(x_{i}-y_{i})^2},
\end{equation}
where $N$ is the population size. Then, the sparsity of the individual $x_{i}$ can be expressed as
\begin{equation}
    SP(x_i)=Z_i/N,
\end{equation}
where $Z_i$ is the number of individuals whose Euclidean distance between it and other individuals is less than the judgment threshold \cite{luan2023enhanced}. Therefore, the individual with the smallest sparsity $SP(x_i)$ is defined as the sparse solution.

After the sparse solution is determined, the children population of local search can be obtained by performing the following two local search operators $O_1$ and $O_2$:
\begin{itemize}
  \item[$O_1$:] Randomly select two elements in the sparse solution and perform the position exchange on them.
  \item[$O_2$:] Randomly pick up one element from the sparse solution and reduce its value by $u$ (e.g., 3).
\end{itemize}

In this context, the sparsity calculation assists the GA in effectively identifying the area that requires more focused local searching, e.g., the sparse solution. Afterwards, the local search operators are performed based on the sparse solution in a more targeted and efficient manner.

\begin{figure}[t]
\vspace{-0.1cm}
\centering
\includegraphics[width=0.45\textwidth]{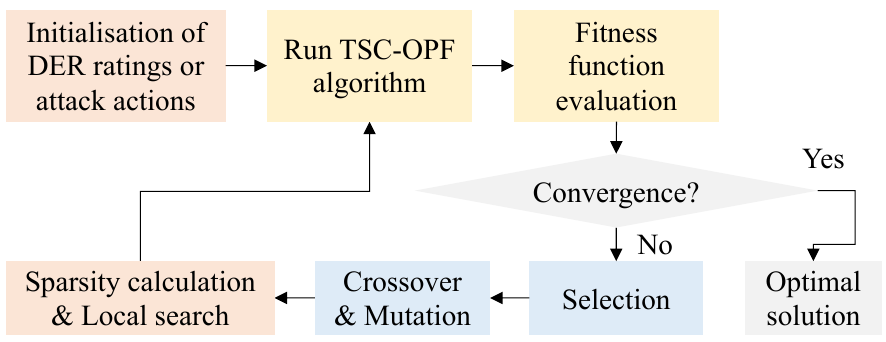}
\vspace{-0.1cm}
\caption{Illustration of the proposed enhanced GA with adaptive crossover and mutation probabilities as well as local search based on sparsity calculation.}\label{fig:aga}
\vspace{-0.35cm}
\end{figure}

\subsubsection{GA Population Design}
As depicted in Fig. \ref{fig:aga}, capacities of DERs for the upper-level problem can be initialised as iterated generations (e.g. pairs of DG ratings [200 kW, 200 kW] with the range of [0 kW, 350 kW]) in the AGA. The fitness function is the total system cost provided iteratively for each chromosome through the TSC-OPF. GA is considered to have converged, if the optimal solution remains unchanged for a specified number of iterations (e.g., 20 or 30) or the maximum allowable number of iterations is reached (e.g., 100). Regarding the low-level problem, various contingency scenarios (e.g., 3000) can be randomly generated, including sudden wind drop, load change, IC outage happening at different time points, etc. As such, one chromosome has to consider two different types of genes in order, e.g., contingency sets and occurrence time points. For example, a chromosome [2,4,1,2,7] represents contingencies $2,4$ occurring at time points $1,2$ of the representative day $7$.

\section{Case Studies}
\vspace{-0.08em}
\label{sec:V}
\subsection{Experimental Setup}
\vspace{-0.00em}
\label{sec:V.A}
\subsubsection{Network and Data Descriptions}
\label{sec:V.A.1}
\begin{figure}[h!]
\vspace{-0.38em}
\centering
{\includegraphics[width=0.48\textwidth]{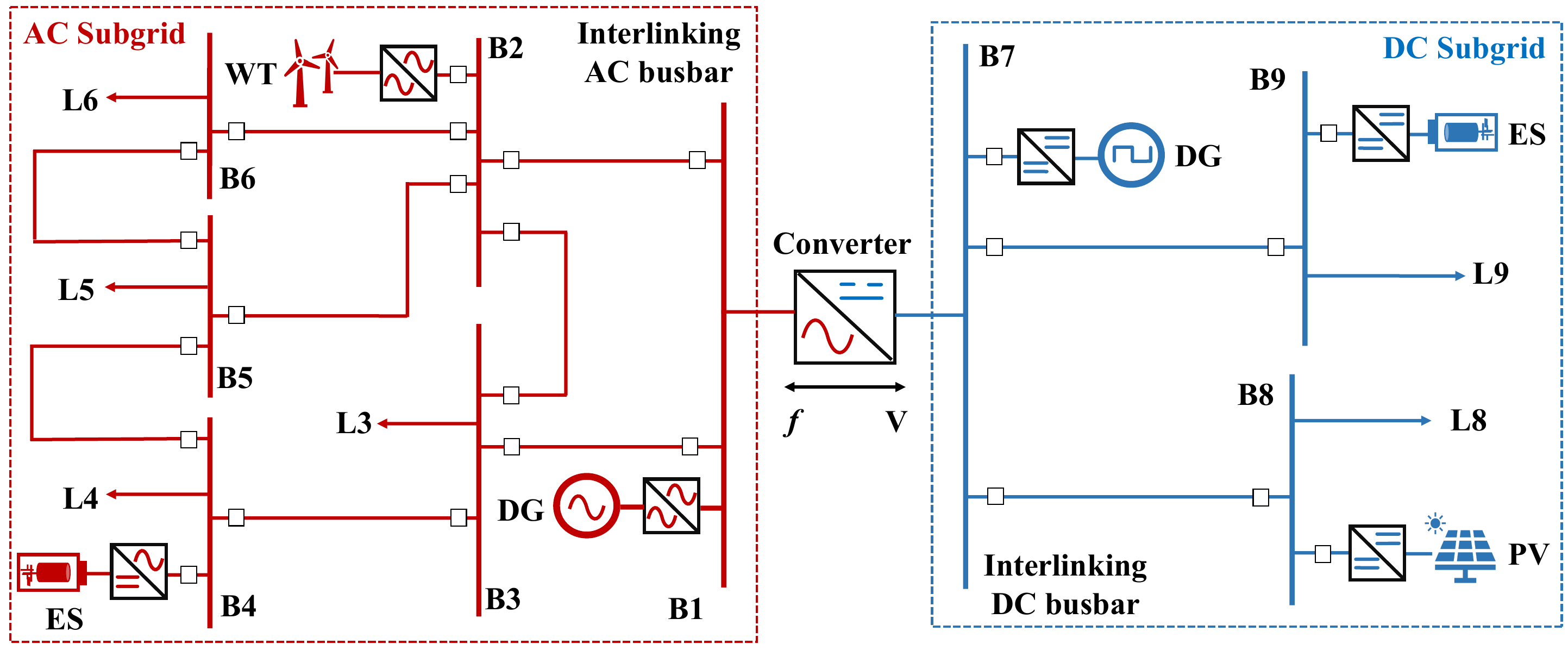}}
\vspace{-0.08em}
\caption{Network structure of the studied AC/DC hybrid MG.}
\label{fig:network}
\vspace{-0.98em}
\end{figure}

Case studies are conducted on an HMG with the network structure illustrated in Fig. \ref{fig:network}. Conventional DGs, WTs, PVs, and ESs are appropriately allocated on the AC and DC sides, while two subgrids are linked via an IC. A real-world yearly dataset collected from the Ausgrid \cite{ratnam2017residential} is used to generate representative days for the planning of the HMG, following the same practices in \cite{nakiganda2022stochastic,rousis2019planning,wang2021three}. The planning period is one year, which is represented by 6 typical days \cite{nakiganda2022stochastic}. Investment cost and operation data for conventional DGs and ESs are collected from \cite{wang2021three}, while the cost and operation data of WTs and PVs can be found in \cite{rousis2019planning}. In detail, important cost parameters of different generation technologies and line data of the studied HMG are summarised in Table \ref{tab:cost_parameters} and Table \ref{tab:line_parameters}, respectively. Finally, to ensure the convergence of the proposed DAD model, the convergence tolerance $\epsilon$ is set to be 0.01\% \cite{zhang2020multi}.

\begin{table}[h!]
\vspace{-0.18em}
\footnotesize
\centering
\renewcommand\arraystretch{1.00}
\setlength{\abovecaptionskip}{4pt}
\caption{Cost Parameters of different generation technologies.}
\setlength{\tabcolsep}{1.98mm}{
\begin{tabular}{ |c|c|c|c|c| }
 	  \toprule
        \begin{tabular}[c]{@{}c@{}}Type \end{tabular}
         & \begin{tabular}[c]{@{}c@{}} Capital\\cost (\pounds /kW)\end{tabular}
 	  & \begin{tabular}[c]{@{}c@{}} Fixed O\&M\\ (\pounds/kW/year)\end{tabular}
 	  & \begin{tabular}[c]{@{}c@{}} Discount \\rate\end{tabular}
        & \begin{tabular}[c]{@{}c@{}} Lifetime \\(year) 
        \end{tabular} \\ \midrule 
        AC-side DG  & 2150 & 41.6 & 9.2\% & 25\\  \midrule 
        DC-side DG  & 760 & 34.4 & 7.8\% & 25\\ \midrule 
        WT  & 1642 & 30.9 & 8.9\% & 23\\ \midrule 
        PV  & 670 & 6.2 & 6.5\% & 25\\ \midrule
        ES  & 3400 & 39.0 & 8.6\% & 20\\ \bottomrule      
\end{tabular}}
\label{tab:cost_parameters}
\vspace{-0.3em}
\end{table}

\begin{table}[h!]
\vspace{-0.88em}
\footnotesize
\centering
\renewcommand\arraystretch{1.00}
\setlength{\abovecaptionskip}{4pt}
\caption{Line data corresponding to the studied AC/DC hybrid MG.}
\setlength{\tabcolsep}{1.98mm}{
\begin{tabular}{|c|c|c|c|}
\toprule
\begin{tabular}[c]{@{}l@{}}Line between \\ bus  $i\rightarrow j$\end{tabular} & \begin{tabular}[c]{@{}l@{}}Reactance \\ ($X_{ij}[p.u.]$)\end{tabular} & \begin{tabular}[c]{@{}l@{}}Resistance \\ ($R_{ij}[p.u.]$)\end{tabular} & \begin{tabular}[c]{@{}l@{}}Maximum \\ flow {[}kVA{]}\end{tabular} \\ \midrule 
1 $\rightarrow$ 2     & 0.20  & 0.10    & 100  \\ 
1 $\rightarrow$ 3    & 0.20   & 0.05   & 100  \\
2 $\rightarrow$ 3   & 0.25  & 0.05    & 60  \\
3 $\rightarrow$ 4   & 0.10  & 0.05   & 60  \\
2 $\rightarrow$ 5   & 0.30  & 0.10  & 60  \\
2 $\rightarrow$ 6   & 0.20  & 0.07  & 60  \\
4 $\rightarrow$ 5   & 0.40 & 0.20 & 60   \\
5 $\rightarrow$ 6   & 0.30 & 0.10 & 60 \\
7 $\rightarrow$ 8   & -    & 0.10    & 120   \\
7 $\rightarrow$ 9   & -  & 0.20  & 120  \\ \bottomrule    
\end{tabular}}
\label{tab:line_parameters}
\vspace{-0.0em}
\end{table}

\subsubsection{Transient Security Setups}
\label{sec:V.A.2}
Regarding frequency limits, the frequency nadir must never fall below 49.2 Hz to avoid triggering under-frequency load shedding \cite{zhao2020frequency}. The RoCoF should remain below 1 Hz/s to prevent the tripping of RoCoF-sensitive protection relays. Additionally, voltage levels should be maintained within the range of 0.9 p.u. to 1.1 p.u. \cite{rousis2019planning,wang2021three}. Furthermore, potential fault scenarios investigated in this paper include sudden wind drop, sudden load change, MG islanding, and IC outage, which are generated via the enhanced GA algorithm suggested in Section \ref{sec:IV.C} and may occur at different time periods to simulate real-world practices. Consequently, the TSC-OPF will be used to capture second-scale frequency and voltage deviations under different fault scenarios.
\vspace{-0.3em}

\subsection{Optimal Sizing Results of DERs}
\begin{figure}[h!]
\vspace{-0.78em}
\centering
{\includegraphics[width=0.49\textwidth]{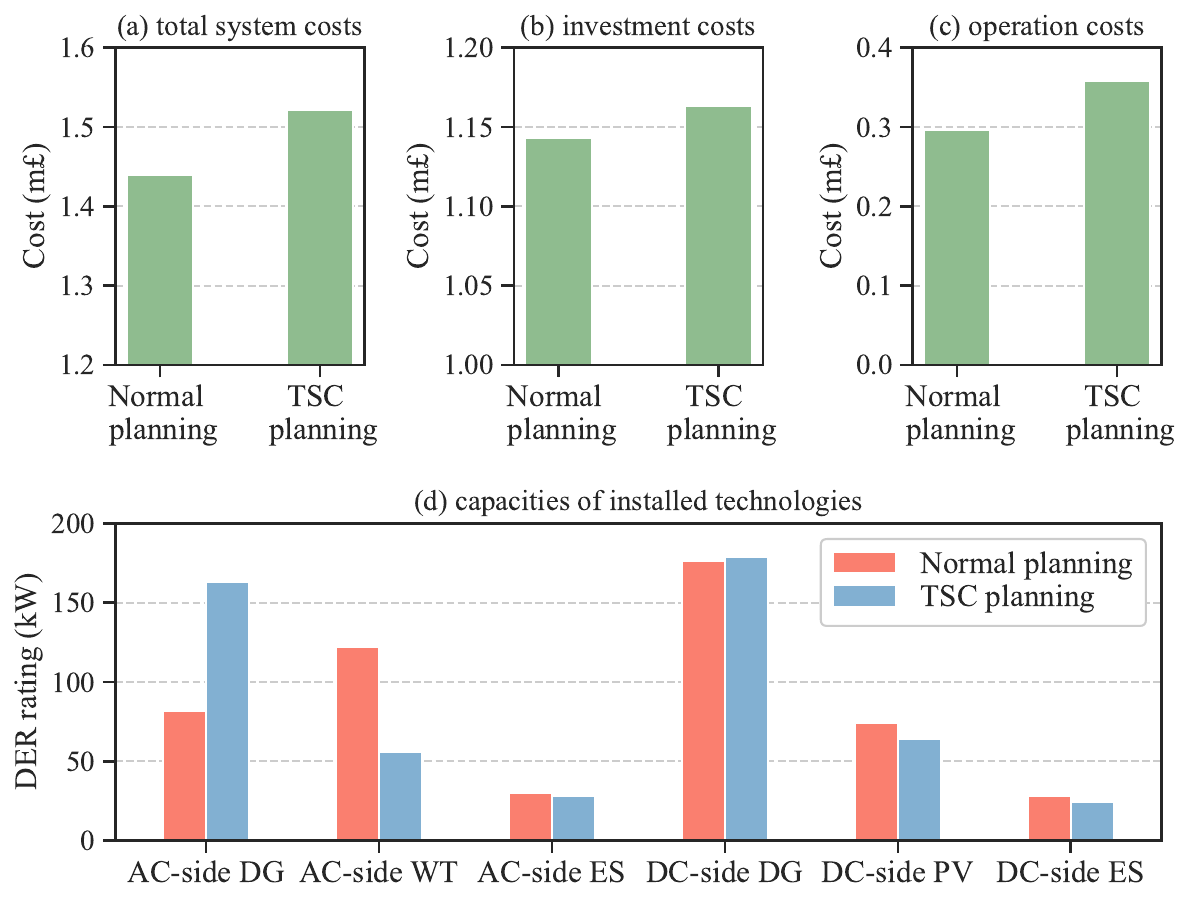}}
\vspace{-1.68em}
\caption{Cost comparison between two scenarios and corresponding installed DER capacities including DGs, WTs, PVs, and ESs.}
\label{fig:nor_tra}
\vspace{-0.20em}
\end{figure}

\vspace{-0.08em}
\label{sec:V.B}
To verify the performance of the proposed transient-stability-driven optimal sizing model, a comparison case study has been conducted, including two scenarios: a) the proposed TSC-OPF model is deployed at the operation stage to capture frequency and voltage dynamics; b) a static hybrid AC/DC power flow model without transient stability considerations is employed for HMG operation. The whole system costs, investment costs, and operation costs under the above two scenarios are respectively depicted in Fig. \ref{fig:nor_tra} (a)-(c), while their optimal sizing results for DERs are illustrated in Fig. \ref{fig:nor_tra}(d).

It can be observed in Fig. \ref{fig:nor_tra} (a)-(c) that considering transient stability constraints in the proposed planning model can lead to higher investment costs and operation costs. This is because the stability-driven planning model aims to ensure stable system operations under worst-case scenarios rather than only seeking low investment costs. Thus, the capacities of conventional DGs have to increase to meet the requirements for frequency and voltage stability, e.g., ensuring enough inertia and frequency reserve to mitigate power imbalance, which leads to higher investment costs and operation costs. On the other hand, as depicted in Fig. \ref{fig:nor_tra}(d), the capacities of WTs and PVs have been significantly reduced, which can mitigate the influence of sudden weather condition changes (e.g., wind speed) on power flows and frequency deviations. In response to smaller RES penetration under stability-driven planning, the capacities of ESs on both AC-side and DC-side have been slightly reduced, due to the high investment costs.

\begin{figure}[h!]
\vspace{-0.38em}
\centering
{\includegraphics[width=0.49\textwidth]{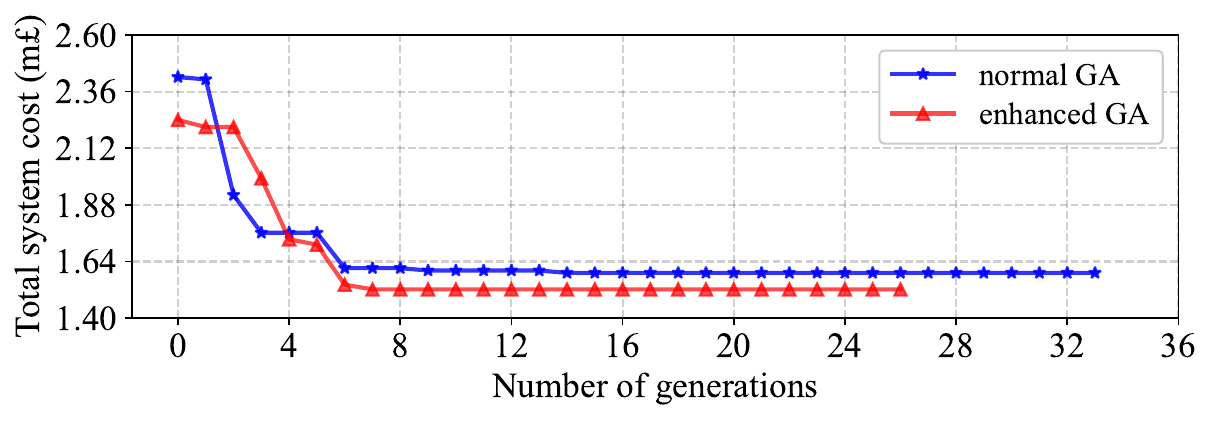}}
\vspace{-1.68em}
\caption{Total system cost evolution curves of the normal GA and the enhanced GA with sparsity calculation and local search.}
\label{fig:aga_result}
\vspace{-0.38em}
\end{figure}

Finally, to demonstrate the performance of the proposed enhanced GA in avoiding local minimum and finding cost-effective solutions, a comparison has been conducted, including two cases: a) normal GA without adaptive probabilities and local search, b) the proposed adaptive GA with sparsity calculation and local search. It can be observed from Fig. \ref{fig:aga_result} that the proposed GA reaches convergence after 27 iterations, which is much faster than the normal GA with 34 iterations. Furthermore, the proposed GA obtains a final solution with a lower system cost than the normal GA, exhibiting its effectiveness in avoiding local minimum.
\vspace{-0.88em}

\subsection{Performance Analysis of the Proposed TSC-OPF}
\begin{figure}[t!]
\vspace{-0.88em}
\centering
{\includegraphics[width=0.49\textwidth]{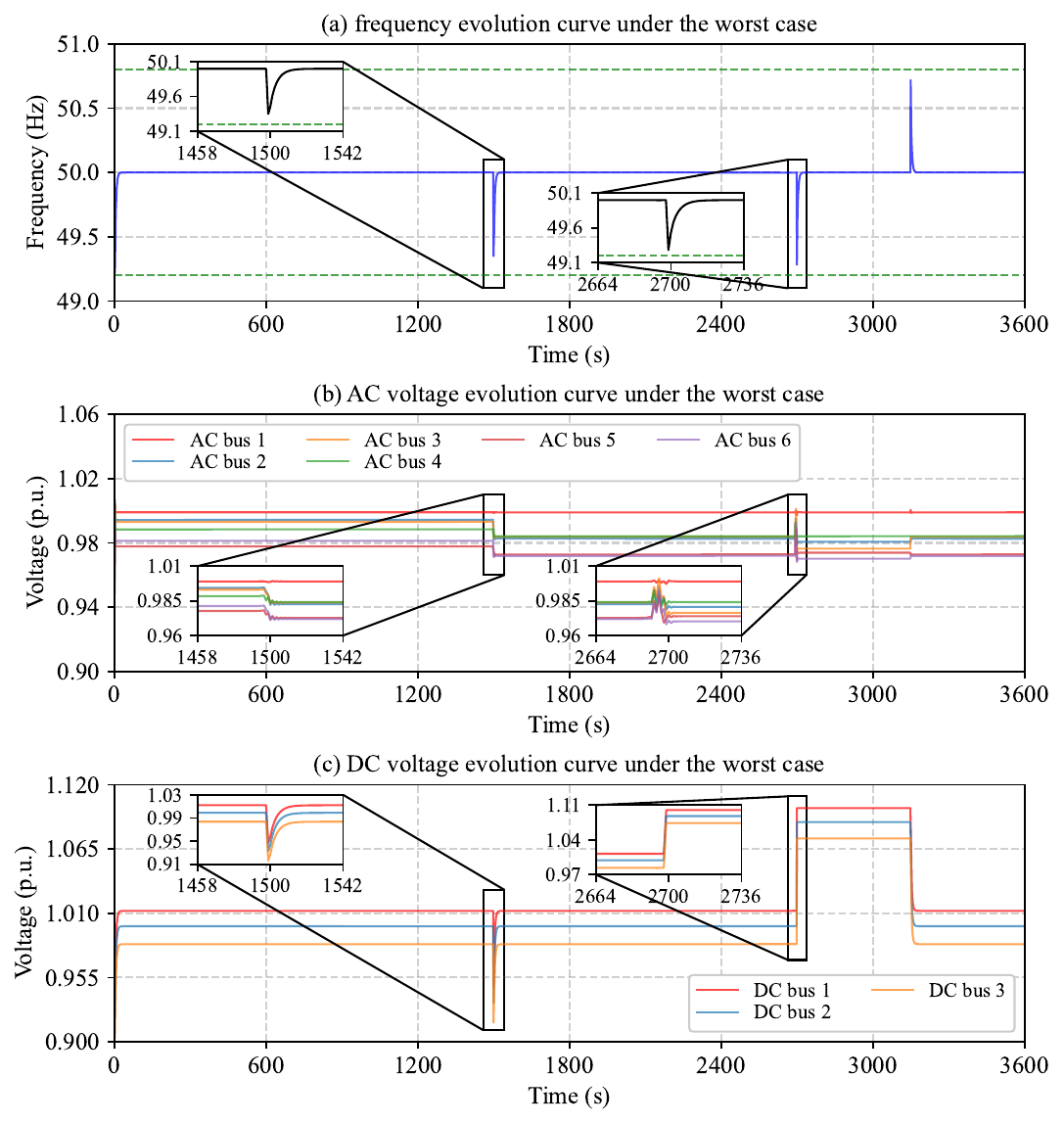}}
\vspace{-1.78em}
\caption{Frequency and voltage profiles under the worst case scenario occurring between 19:00 and 20:00 of the selected representative day.}
\label{fig:f_v_tran}
\vspace{-1.28em}
\end{figure}

\begin{figure}[h!]
\vspace{-0.88em}
\centering
{\includegraphics[width=0.49\textwidth]{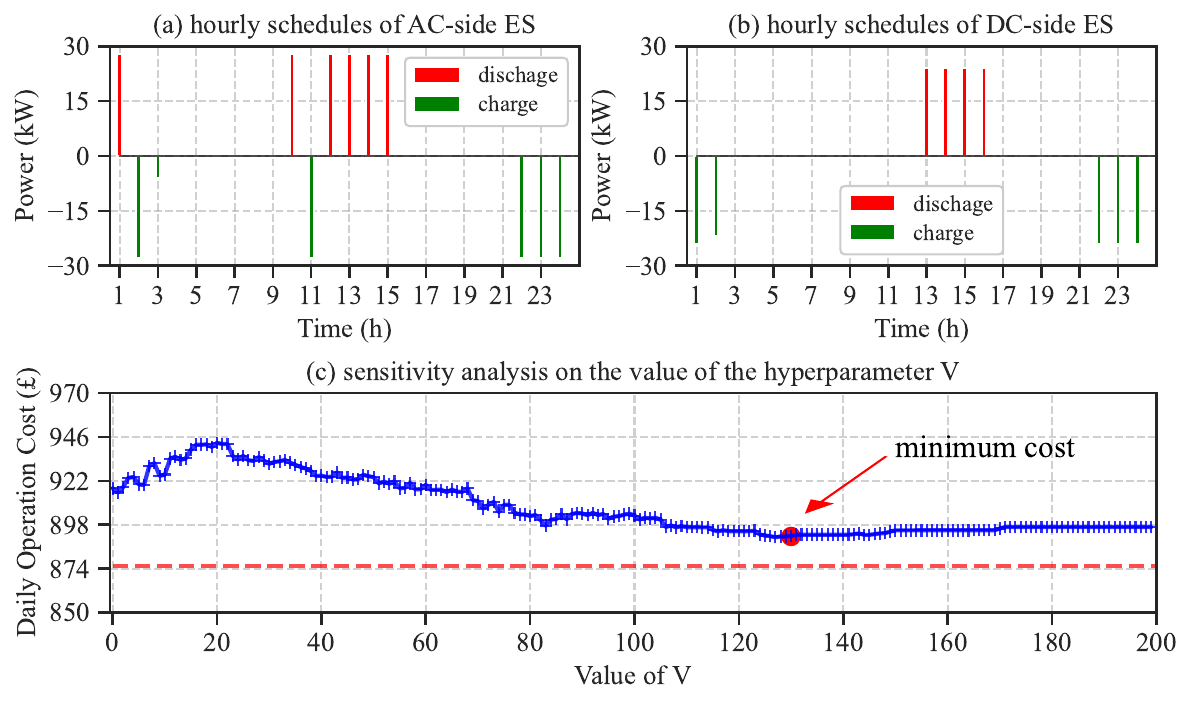}}
{\includegraphics[width=0.49\textwidth]{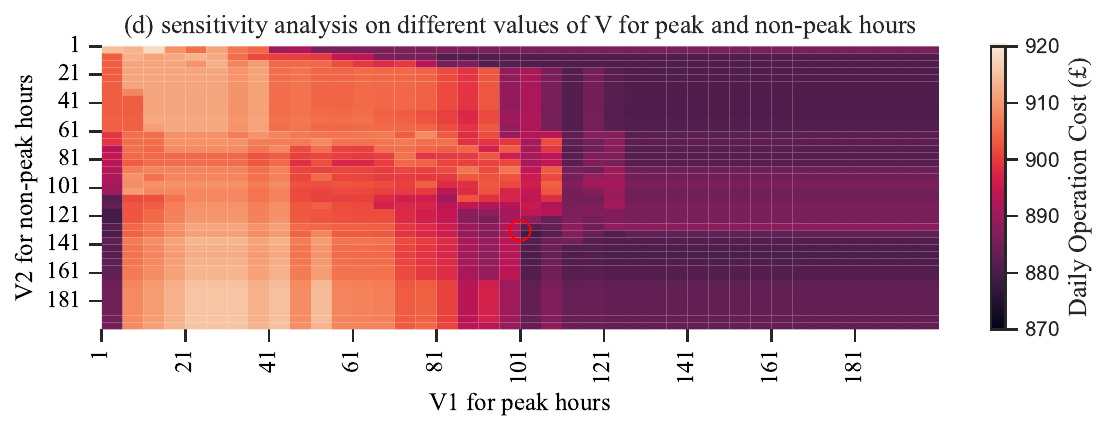}}
\vspace{-1.78em}
\caption{Illustration of scheduling results of Lyapunov optimisation on the selected representative day as well as sensitivity analysis on the value of $V$.}
\label{fig:lo_re}
\vspace{-0.58em}
\end{figure}

\vspace{-0.08em}
\label{sec:V.C}
This section serves as a verification to analyse the performance of the proposed operation model including TSC-OPF and Lyapunov optimisation. Specifically, the transient stability dynamics of frequency and voltage under the worst-case scenario occurring at hour 19 (between 19:00-20:00) of one representative day are selected and exhibited in Fig. \ref{fig:f_v_tran}. The power charging and discharging behaviours of ESs under the proposed Lyapunov optimisation framework ($V=100$) during this selected day are depicted in Fig. \ref{fig:lo_re} (a)-(b), while the results of sensitivity analysis on the value of $V$ and multiple $V$s are illustrated in Fig. \ref{fig:lo_re} (c)-(d), respectively.

It can be observed from Fig. \ref{fig:f_v_tran} that there are two contingencies during the selected hour (19:00-20:00), contributing to the worst-case scenario together. In detail, the contingency between 1,458 sec. and 1,542 sec. corresponds to a sudden big wind drop, while the contingency between 2,664 sec. and 2,736 sec. is caused by the sudden IC outage. Fig. \ref{fig:f_v_tran}(a) shows that the proposed TSC-OPF can accurately capture the frequency deviations during these two contingencies, and the optimal sizing decisions for DERs can successfully rescue the frequency curve before it reaches the Nadir (49.2 Hz). Accordingly, there are significant voltage fluctuations in both AC and DC sides when these two contingencies occur. Specifically, it can be found from Fig. \ref{fig:f_v_tran}(b) that the voltage level at AC bus 1 is stable and maintained at the reference value (1.00 p.u.) due to the incorporation of AVR, further verifying the performance of the proposed TSC-OPF on frequency and voltage control. In addition, the voltage profiles on the DC side drop when the first contingency occurs, following the trend of frequency deviation to maintain stable power exchange between the AC and DC sides. On the other hand, DC voltage profiles increase and reach the maximum value during the second contingency, since there is no power exchange or linking between AC and DC sides when the IC is outaged.

Regarding the performance of the proposed Lyapunov optimisation approach, it can be observed from Fig. \ref{fig:lo_re} (a)-(b) that ESs on both AC and DC sides can make realistic power charging and discharging behaviours during the selected day. Furthermore, the influence of hyperparameter $V$ on daily operation costs is illustrated in Fig. \ref{fig:lo_re}(c), where the red dashed line represents costs for the optimal solution of $875.6\pounds$ under a time-coupled optimisation. It can be found that the operation costs from Lyapunov optimisation are close to the optimal cost, where a minimum cost of $891.6 \pounds$ is achieved when the value of $V$ is 130. In general, selecting a $V$ that is larger than 110 can always ensure a good optimisation performance, verifying the stability of the proposed Lypapuov optimisation method.

Additionally, a further sensitivity analysis is conducted in Fig. \ref{fig:lo_re}(d) to investigate the performance of using two different $V$s for peak and non-peak hours, respectively. It can be found that using two different $V$s indeed adds further flexibility to the proposed Lyapunov optimisation approach and obtains a minimum cost of $880.1 \pounds$ when $V_1\!=\!105$ and $V_2\!=\!135$, which is closer to the optimal cost $875.6\pounds$ compared with the single-$V$ Lyapunov optimisation. However, it is worth noting that the proposed Lyapunov optimisation approach is primarily used to capture a trade-off between operation costs and the depth of discharge of ESs rather than only seeking the minimum costs, which is also aligned with the usage requirements of ESs in a planning problem, since frequent charging and discharging behaviours may compromise the service life of ESs. 
\vspace{-0.68em}

\subsection{Simulating $N\!\!-\!K$ Contingencies}
\label{sec:V.D}
\begin{figure}[t!]
\vspace{-0.48em}
\centering
{\includegraphics[width=0.49\textwidth]{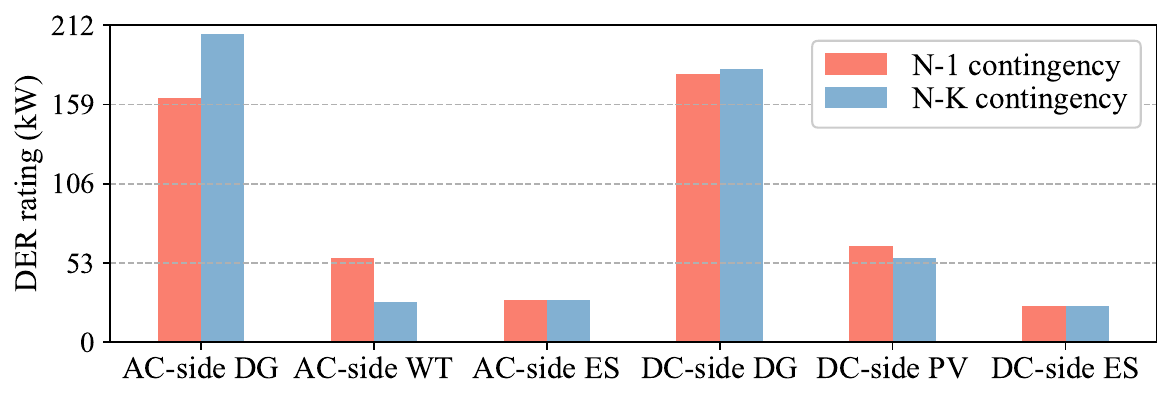}}
\vspace{-1.68em}
\caption{Comparison of optimal sizing results under two different cases, i.e., $N-1$ and $N-K$ contingencies.}
\label{fig:double}
\vspace{-1.00em}
\end{figure}

In this section, we assume that two contingencies may occur simultaneously to simulate potential $N\!-\!K$ scenarios, while at most three different contingencies can happen in one time period. Both the optimal sizing results under $N-1$ contingency and $N-K$ contingency are demonstrated in Fig. \ref{fig:double} for comparison. It can be found that the capacity of AC-side DG under $N-K$ contingency has increased significantly, compared with its capacity under $N-1$ contingency. This is because much more inertia and frequency response are required to maintain frequency stability when two contingencies occur simultaneously and cause larger power infeed loss. Accordingly, the capacity of AC-side WT has been reduced under $N-K$ contingency. Except for the complementary effect with AC-side DG, reduced WT capacity can mitigate the influence of potential wind drop on frequency deviations.

\begin{figure}[h!]
\vspace{-0.38em}
\centering
{\includegraphics[width=0.49\textwidth]{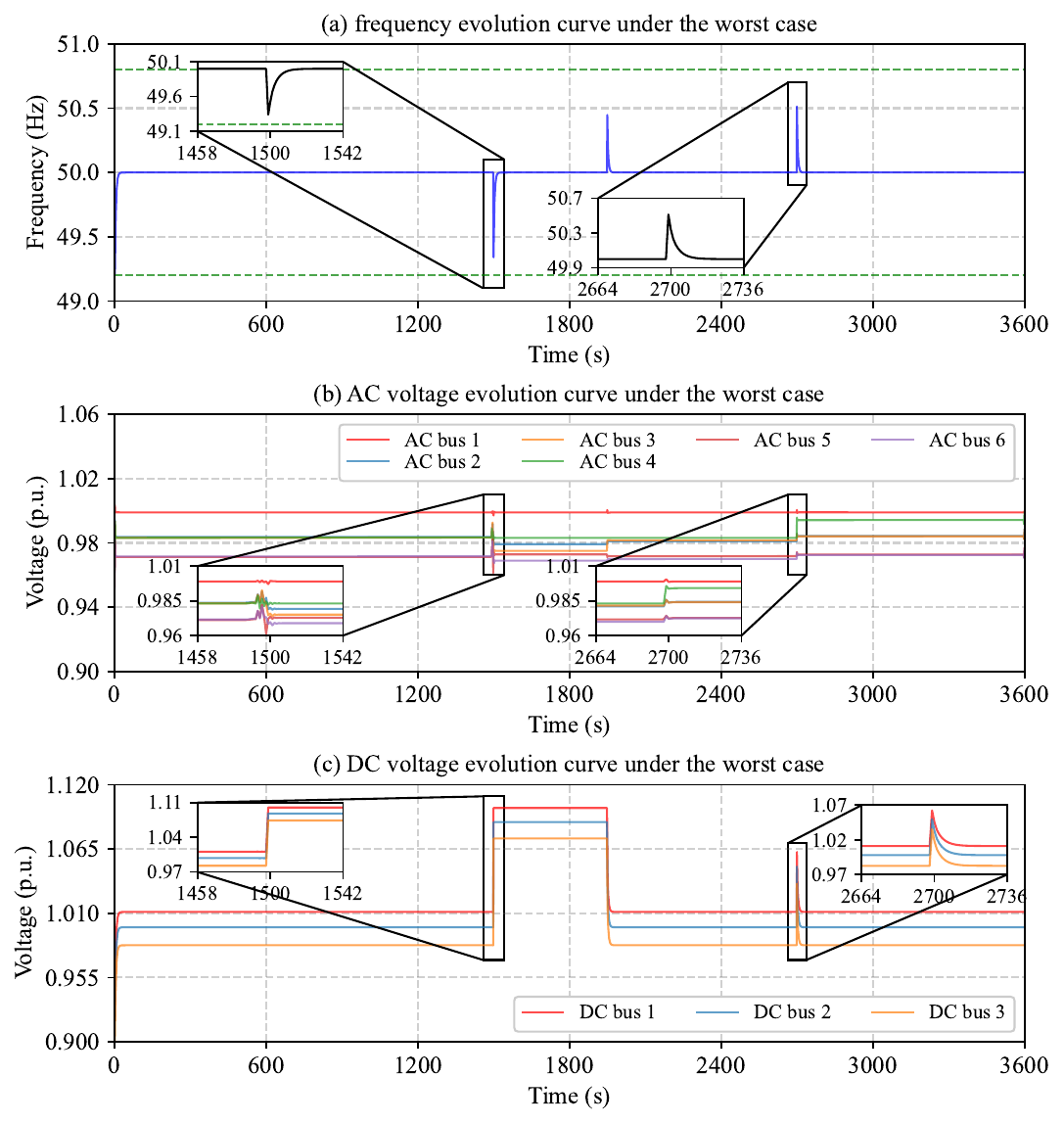}}
\vspace{-1.78em}
\caption{Frequency and voltage profiles under the worst-case scenario where two contingencies occur simultaneously.}
\label{fig:f_v_2}
\vspace{-0.38em}
\end{figure}

Regarding the performance of the proposed TSC-OPF, Fig. \ref{fig:f_v_2} (a)-(c) depict the frequency and voltage deviations under the worst-case scenario, where two contingencies including large wind drop and IC outage occur simultaneously between 1458 sec. and 1542 sec. It can be found that the frequency drop has been rescued before reaching the Nadir under the sizing results, while AC voltage profiles are restricted within the allowable range due to the consideration of AVR and droop control schemes. Thus, the performance of the proposed stability-driven planning framework has been verified again. 
\vspace{-0.7em}

\subsection{Optimal Sizing for both DERs and ICs}
\begin{figure}[h!]
\vspace{-0.28em}
\centering
{\includegraphics[width=0.49\textwidth]{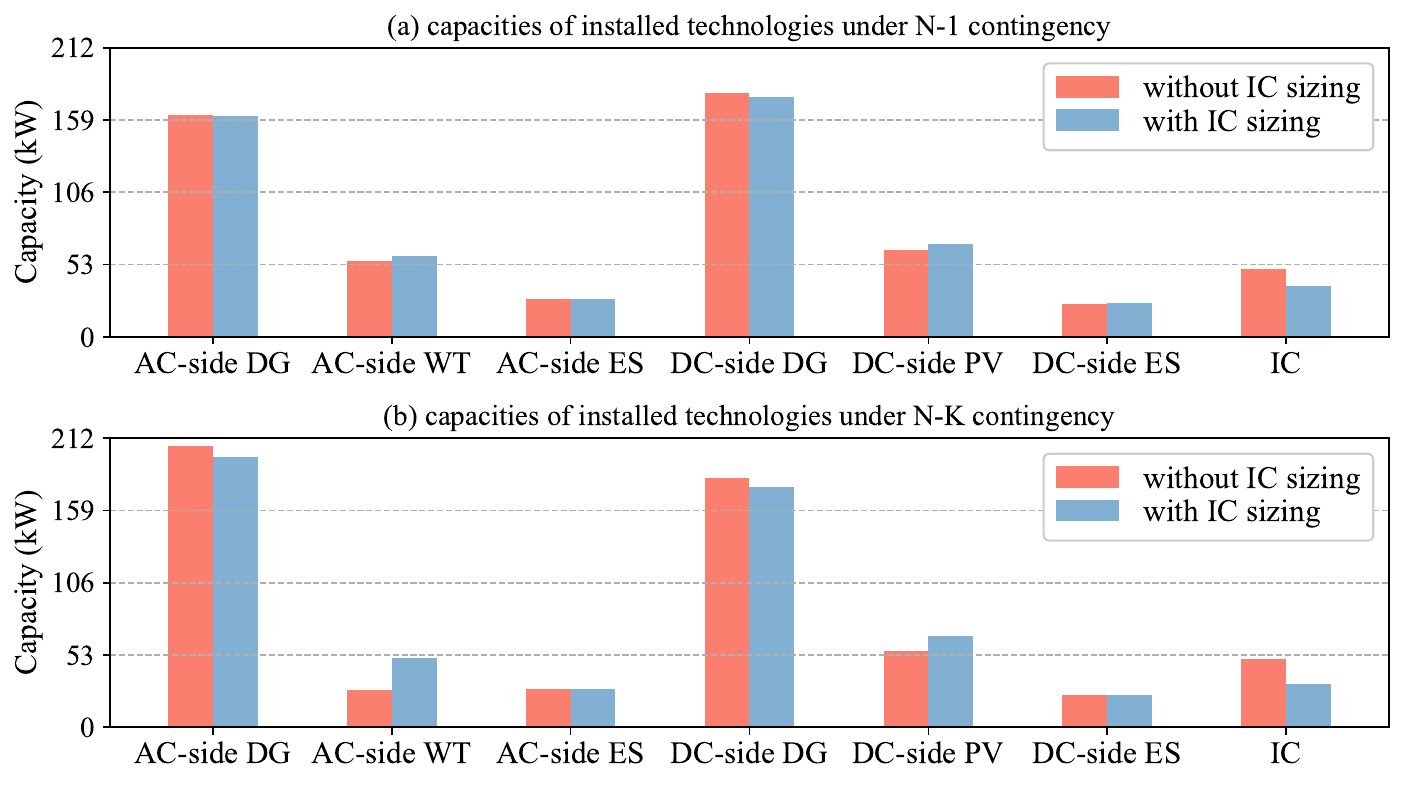}}
\vspace{-1.08em}
\caption{Comparisons of installed capacities between two cases `without IC sizing' and `with IC sizing'.}
\label{fig:big_cost}
\vspace{-0.1em}
\end{figure}

\vspace{-0.0em}
\label{sec:V.E}
This section is used as a further illustration of the proposed planning approach on scalability through the optimal sizing of both DERs and ICs. With respect to the optimal sizing results without IC sizing provided in Section \ref{sec:V.B} and Section \ref{sec:V.D}, the investment results with IC sizing are illustrated in Fig. \ref{fig:big_cost}. 

It can be observed from Fig. \ref{fig:big_cost}(a) that the sizing results of DERs under the case `with IC sizing' are similar to the case `without IC sizing', while the IC capacity is reduced from 50 kW to 37 kW when IC is involved in the planning framework. It can be seen that there exists significant over-investment if the IC is not considered in the planning framework. On the other hand, under the scenario $N-K$ contingency, it can be found that the IC capacity has been significantly reduced from 50 kW to 33 kW, which can mitigate the influence of potential IC outages on frequency and voltage deviations. Accordingly, the capacity of AC-side DG has been slightly reduced, while the capacities of AC-side WT and DC-side PV increase due to complementary effects. Overall, it can be concluded that the planning framework can provide realistic sizing results for different types of resources, further verifying its effectiveness.
\vspace{-1.08em}

Finally, the computational time of the proposed method is reported in Table \ref{tab:computing}, compared with two benchmarks: 1) normal planning without transient stability considerations; 2) stability-driven planning with a normal GA, i.e., no adaptive probabilities and local search capabilities. All the case studies were run on Intel Core i7-8700 processor with 16 GB memory.

\begin{table}[h!]
\vspace{-0.98em}
\footnotesize
\centering
\renewcommand\arraystretch{1.00}
\setlength{\abovecaptionskip}{4pt}
\caption{Computational performance of three different methods.}
\setlength{\tabcolsep}{2.78mm}{
\begin{tabular}{ |c|c|c|}
 	  \toprule
        No.
        &\begin{tabular}[c]{@{}c@{}}Method \end{tabular}
         & \begin{tabular}[c]{@{}c@{}} Computation time (hour)\end{tabular}\\ \midrule 
        1 & \begin{tabular}[c]{@{}c@{}}Normal planning model \\ without transient stability\end{tabular}  & 1.02 \\ \midrule 
        2 & \begin{tabular}[c]{@{}c@{}}Stability-driven planning \\ with a normal GA\end{tabular}  & 5.44 \\ \midrule 
        3 & \begin{tabular}[c]{@{}c@{}}The proposed method, i.e., \\stability-driven planning \\ with the enhanced GA\end{tabular}  &  4.30\\ \bottomrule      
\end{tabular}}
\label{tab:computing}
\vspace{-0.4em}
\end{table}

It can be observed from Table \ref{tab:computing} that the normal planning model obtained the shortest computing time (1.02 hours) due to the lack of transient stability considerations, while the stability-driven planning model with a normal GA received the longest computing time (5.44 hours) because of the incorporation of DAEs. After adopting the enhanced GA with sparsity calculation and local search as well as adaptive crossover and mutation probabilities, the computing time has been significantly reduced to 4.30 hours, compared with the normal GA. Therefore, the performance of the enhanced GA on avoiding local minimum and reducing computing time has been demonstrated thoroughly. Finally, it is worth noting that the key focus of the proposed planning model is to obtain accurate planning decisions that can ensure frequency and voltage stability against various contingencies with the minimum cost rather than reducing computing time.
\vspace{-0.5em}

\section{Conclusions}
\label{sec:VI}
In this paper, a transient stability-driven planning approach is proposed to solve the optimal sizing problem of HMGs capturing frequency and voltage stability. The proposed planning framework is formulated as a DAD model with a two-level structure, where an enhanced AGA algorithm with sparsity calculation and local search is developed to solve both low-level and upper-level problems. A novel TSC-OPF model integrated with Lyapunov drift-plus-penalty optimisation is proposed and deployed at the operation level to consider both static and transient dynamics with detailed frequency and voltage control loops for DGs and IBRs as well as IC frequency-voltage coupling characteristics. Experiments are conducted on an HMG system with ICs to evaluate the superiority of the proposed stability-driven planning approach in achieving realistic sizing decisions for DERs that can maintain frequency and voltage stability during severe transient contingencies. Both $N-1$ and $N-K$ contingencies are simulated and compared, while the algorithm scalability is verified by considering the optimal sizing problem of both DERs and ICs. It can be found that the proposed TSC-OPF model can successfully capture frequency and voltage deviations under various contingencies and ensure frequency and voltage stability via embedded dynamic control loops. On the other hand, the provided sizing decisions from the enhanced GA can successfully obtain cost-effective investment solutions with stability guarantees. Even though more investment is generally needed for stability concerns compared with normal planning, the reliability and resilience of the HMG system have been significantly enhanced, reducing the risk of economic losses from various instability issues.
\vspace{-0.28em}

\bibliographystyle{IEEEtran}
\bibliography{References.bib}

\begin{IEEEbiography}[{\includegraphics[width=1in,height=1.25in,clip,keepaspectratio]{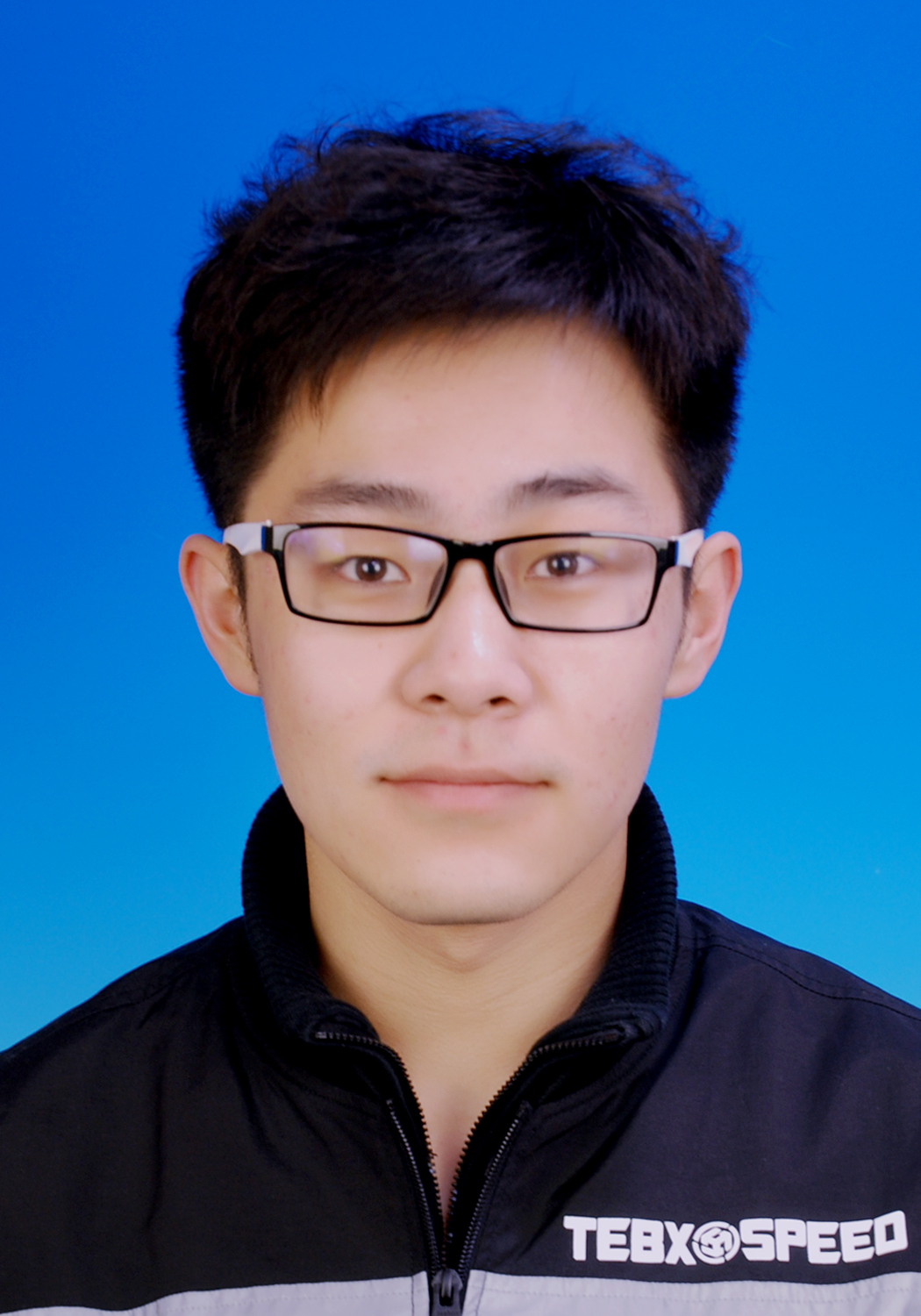}}]{Yi Wang}
(Member, IEEE)
received the B.Eng. degree and the M.Eng. degree from Tianjin University in 2015 and 2018, and the Ph.D. degree from Imperial College London in 2022. He is currently employed as a Research Associate in the Department of Electrical and Electronic Engineering at Imperial College London. His research interests include mathematical programming and learning approaches applied to the planning and operation of networked microgrids, the resilience enhancement of future power systems, frequency-constrained power system optimisation, and multi-energy system integration.
\end{IEEEbiography}

\begin{IEEEbiography}[{\includegraphics[width=1in,height=1.25in,clip,keepaspectratio]{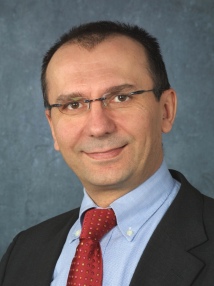}}]{Goran Strbac}
(Member, IEEE) 
is a Professor of Energy Systems at Imperial College London, London, U.K. He led the development of novel advanced analysis approaches and methodologies that have been extensively used to inform industry, governments, and regulatory bodies about the role and value of emerging new technologies and systems in supporting cost effective evolution to smart low carbon future. He is currently the Director of the joint Imperial-Tsinghua Research Centre on Intelligent Power and Energy Systems, Leading Author in IPCC WG 3, Member of the European Technology and Innovation Platform for Smart Networks for the Energy Transition, and Member of the Joint EU Programme in Energy Systems Integration of the European Energy Research Alliance.
\end{IEEEbiography}

\end{document}